\documentclass[sn-aps]{sn-jnl}% American Physical Society (APS) Reference Style
\usepackage{braket}
\usepackage{epsfig}
\usepackage{bm}
\DeclareUnicodeCharacter{030C}{\'{c}}

\begin{document}

\title{Giant Rashba electrical control of magnetism in band models
}

\author*{\fnm{Wen} \sur{Li}}\email{wxl386@miami.edu}
\author*{\fnm{Stewart} \sur{Barnes}}\email{sbarnes@miami.edu}

%\author{ Wen Li and Stewart E.  Barnes}

\affil{\orgdiv{Physics Department}, \orgname{University of Miami}, \orgaddress{%\street{Street}, 
\city{Coral Gables}, \postcode{33124}, \state{FL}, \country{USA}}}

%\affiliation{Physics Department, University of Miami, Coral Gables, FL 33124, USA. }

\date{\today}

%\bibitem{Wang2020}  See recent review: L. Z. Wang, X. Li, T. Sasaki, K. Wong, G. Q. Yu, S. Z. Peng, C. Zhao, T. Ohkubo, K. Hono, W. S. Zhao, and K. L. Wang, High voltage-controlled magnetic anisotropy and interface magnetoelectric effect in sputtered multilayers annealed at high temperatures, Sci. China-Phys. Mech. Astron. 63, 277512 (2020), https://doi.org/10.1007/s11433-019-1524-y

%Veit, M.J., Arras, R., Ramshaw, B.J. et al. Nonzero Berry phase in quantum oscillations from giant Rashba-type spin splitting in LaTiO3/SrTiO3 heterostructures. Nat Commun 9, 1458 (2018). https://doi.org/10.1038/s41467-018-04014-0

\abstract{
It is of considerable technological importance to achieve an electrical control of magnetism of sufficient magnitude.
To overcome the in-plane shape anisotropy, needed is the electrical control of a perpendicular magnetic anisotropy (PMA).  It is known, 
 within a free electron model, the Rashba spin-orbit coupling provides such a control. Surprisingly, this same Rashba PMA is enhanced by two to three orders of magnitude when a  periodic potential is added. Usually spin Berry phase physics reflects time dependent magnetic fields. 
 Here it is shown, within a time independent model, such physics arises because the Rashba effective magnetic field has texture within the unit cell.
 Predicted are electrical controllable band-structure gaps, linear in the applied electric field $E$, that can result in a truly giant linear PMA.  Also possible is a Peierls mechanism, in which the magnetisation tilts from the vertical, shifting these gaps to the Fermi level.  As a consequence there are low dissipation electric field driven dynamics, an alternative to the more dissipative spin torque transfer (STT) effect. The theory requires the introduction of an intrinsic spin Berry connection $\vec A_s$, an effective vector potential, and is incompatible with current density functional theories (DFT).
}

\maketitle

For magnetic memory applications, it is estimated\cite{Wang2020}, an electrical control of magnetism  coefficient of  $\gtrsim200$ fJ/Vm is needed. 
This remains an elusive goal. Here the more modest aim is to find an effective Rashba magnetic field $\vec B_R$ of magnitude $\sim 1-10$T. 
 Within the free electron model, Barnes et al\cite{Barnes2014} have shown the Rashba spin-orbit-coupling can generate the necessary PMA, proportional to $E^2$, but for two dimensions, there are two contributions to the magnetic anisotropy that almost cancel. This near cancellation no longer occurs when a  periodic potential $V(\vec r)$ is accounted for. The PMA contribution dominates leading to a massive  increase in this anisotropy energy. It addition,
for  a (near)  perpendicular magnetisation $\vec M$, the periodic potential introduces spin Berry phase physics described by a connection $\vec A_s$ proposed some time ago \cite{Barnes2012}. Even in one dimensions, as an electron is repeatedly reflected,  it precesses on the Bloch spin sphere. The resulting spin mixing  opens electrically controlled band structure gaps,  linear in $E$, near points at which there is an accidental degeneracy of the majority/minority bands. For suitable material parameters, implied is a a truly giant PMA. A Peierls effect\cite{Gruner1988,Barnes1983} is 
another consequence of this spin Berry phase physics.  The magnetisation $\vec M$ tilts away from the vertical thereby shifting the spin gaps to the Fermi level. Implied, in addition to a giant PMA, are electric field driven dynamics, an alternative to the more dissipative spin torque transfer (STT) effect\cite{Slonczewski1996,Barnes2012}. 

Kept in mind, as an illustrative example, are ultra-thin magnetic layers with a PMA induced by an equally thin Au layer. There is a small exchange splitting induced by the magnetic proximity effect. The aim is the have an exchange field $\vec B_E$ of a similar magnitude to the effective Rashba field $\vec B_R$.

{\it Spin-orbit coupling.} While the spin-orbit coupling is implicit in the Dirac equation\cite{Messiah1966}  it appears explicitly, as $V_{\rm so} = - 
\frac{e\hbar }{4m^2 c^2}\vec E \cdot  (\vec \sigma\times \vec p) =  - \mu_B \frac{1}{2c^2}  \vec E \cdot  (\vec \sigma\times \vec v)
= -  \frac{ \lambda_c}{8\pi mc}  \vec E \cdot  (\vec \sigma\times \vec p)$, where $\lambda_c \approx 2$pm is the Compton wave-length, and $\mu_B$ the Bohr magneton, only when the Dirac  equation is reduced to that of Pauli using e.g., the Foldy-Wouthuysen (FW) transformation\cite{FW1950}.  It is customary\cite{Winkler2003} to insist $\vec E$ is large only in the core region where the wave-functions are atomic like and it  follows\cite{Messiah1966}  $V_{\rm so} $ reduces to the more familiar $v_{\rm so} =\xi(r)  \vec \sigma \cdot \vec \ell\, $ where $\xi (r) =\frac{e\hbar }{4m^2 c^2} \frac{1}{r} \frac{d V_1(r)}{dr} $ and where $V_1(r)$ is the Coulomb potential of the atomic core\cite{Winkler2003,Wang1996}. 
It is invariably $v_{\rm so}$ that is used to develop a theory of the Rashba effect\cite{Rashba1984,Winkler2003,Ong2015,Yao2004,Manchon2015,Blaha2001} 
and it is this that is  included in the popular flavours of the DFT\cite{Kresse1993a,Kresse1993b,Kresse1993c,Kresse1993d}. 
A number of methods\cite{Koseki2019}, give accurate values of $v_{\rm so}$ and such information is included 
 in the best DFT. However, it remains useful to  estimate $v_{\rm so}$\cite{Landau1991,SupMat}. 
 Following Landau and Lifshitz\cite{Landau1991}:  
 $\langle \xi (r) \rangle \approx \frac{1}{2}\frac{1}{n(\ell + \frac{1}{2})(\ell+1)} (\alpha Z)^2 \lvert E_{n}\rvert$ where $E_n =- \frac{me^4}{2\hbar^2 } \frac{1}{n^2} $ are the hydrogen energy eigenvalues, $\alpha $ is the fine-structure constant and $Z$ the atomic number. This is consistent with the fundamental idea that the $V_{\rm so} $ is the second order $(\alpha Z)^2, $ QED correction to the energy $E_{n}$ of an outer electron. Usually\cite{Landau1991} mixing with the core 1s-electron is accounted for by setting the principal and angular momentum quantum numbers to $n=1$ and $\ell=0$  resulting in $\langle \xi (r) \rangle \approx 2.6$eV for $Z=79$ appropriate to Au. 
 However that $\vec \ell$  be finite, it is rather mixing with the core 2p-electrons with  $n=2$ and $\ell=1$ that is appropriate. This drastically reduces $\langle \xi (r) \rangle \approx 100$meV, and is actually consistent with the Rashba splitting of the surface state of Au. A similar estimate\cite{radialint,SupMat}  for the Au electric field in the core is $\langle E\rangle \approx  1.2  \times 
10^{16}$V/m while the $\ell =1$ tangential velocity $v \sim 0.15c$. It is hard to imagine a gating field greater than $E \sim 10^{11}$V/m leading to Rashba field $B_R \sim 10^{-2}$T when corrected by the usual  factor of $\frac{1}{Z^2}$\cite{Landau1991,SupMat}.
Such small fields are of little technological importance. 
Here, it is not disputed that the DFT can accurately include $v_{\rm so}$, rather it is observed that $\vec \ell = \vec r \times \vec p$ accounts only for the tangential component of $\vec p$ while for a more or less uniform $\vec E$, due to either a surface, a lack of inversion symmetry, and/or gating, it is the entire in-plane $\vec p$ that counts and account must be taken of the entire 
spin-orbit coupling $V_{\rm so}$. 

Evidently the free electron Rashba model  considered  by Barnes et al.\cite{Barnes2014}  lacks a radial potential $V(r)$ and $v_{\rm so}$ is simply absent. Unfortunately the magnitude of the predicted PMA is many orders of magnitude too small to be interesting, and worse, in two dimensions a term that favours an in plane magnetisation cancels the PMA to leading order\cite{Barnes2014}. In this case the full $V_{\rm so} = \frac{\alpha_R}{\hbar} \vec \sigma\cdot (\vec p\times \hat{\bf z})$, for $\vec E = \hat{\bf z}E$, constitutes the Rashba spin-orbit coupling and importantly involves the canonical momentum $\vec p = \hbar \vec k $ and not just the crystal momentum $\hbar \vec q$. 
The current exercise extends Ref.~\cite{Barnes2014}   by including the effects of a periodic potential in a toy Kronig-Penny model for which $v_{\rm so}$ remains  absent. Bloch's theorem now insists, e.g., a one dimensional  wave-function is a linear combination of $e^{i(q+ G_n)x}$ that corresponds to a canonical momentum $p = \hbar k = \hbar q + \hbar G_n$ that differs from the crystal momentum $\hbar q$ by $\hbar G_n$ where $G_n$ is a reciprocal lattice vector. For the 6$s$-electron surface state  of Au the $\hbar G_n$ are  orders of magnitude larger than $\hbar q$ and relevant vectors corresponds to a velocity  $v = \frac{p}{m} \approx \frac{\hbar G_n}{m} \approx \frac{c}{10}$ that evidently  is not concentrated in the non-existent core. It is implied for Au and a magnetic proximity effect, a PMA several orders larger than that predicted by the free electron model\cite{Barnes2014}. In addition, introduced  are electrically controllable spin gaps directly proportional to $E$, and the periodic  potential $V$, leading to a potential giant PMA. Further discussion of the spin-orbit coupling, along with more details of many of the calculation, is to be found in the Supplementary Materials corresponding to Ref.~\cite{SupMat}.

 \begin{figure}[b]
 \vskip -20pt
\begin{center}
\includegraphics[width=7.5cm]{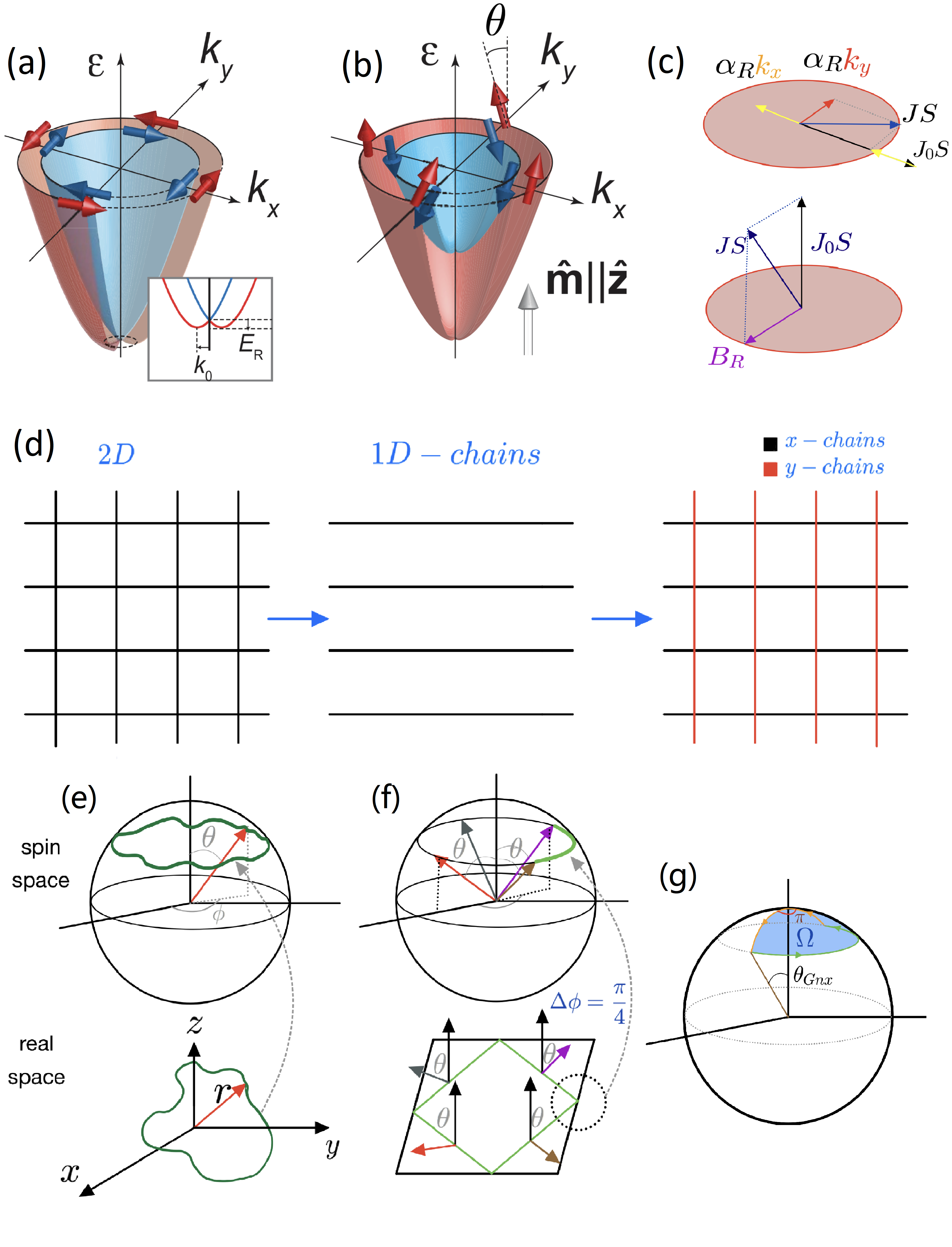}
\end{center}
       \vspace*{-15pt}  
        \caption{ \small (a) With a perpendicular electric field $\vec {E} = E\bm{z}$ and a given momentum $\vec {p}$, the Rashba field $\vec {B}_R$, proportional to $\vec {p} \times \vec {E}$ is tangential to the circles of constant $p$. (b) In the magnetic case $\vec {B}_R$ has, in addition, a perpendicular component and the axis of quantisation of the spin makes and angle $\theta$ to $\vec {E}$. The angle $\theta$ and $\phi$ determine the Berry phase. (c) Top, with an in-plane magnetisation the Rashba field $\vec {B}_R$ is also in-plane and adds to the exchange field $\vec {B}_E$. Bottom, the magnetisation and exchange field are perpendicular to the plane while $\vec {B}_R$ is in-plane and adds vectorially to give an effective exchange that makes a momentum dependent angle  $\theta$ to the perpendicular. 
        (d) Two dimensions corresponds to an array of atoms but can be reduced to a series of crossed chains.  (e) A real space path has an image in spin space, i.e., the Bloch sphere. 
        (f) Upon each  reflection from perpendicular $\delta$-walls, there is a rotation by $\Delta \phi =\pi/2$ about the $\bm{z}$-axis with a constant angle $\theta$. There is a solid angle $\Omega = 2\pi(1-\cos \theta)$ subtended by this path, leading to a Berry phase $\frac{1}{2} \Omega$ and a finite flux. (g) The up-spin is first rotated by the angle $\theta_{G_{n_x}}$ away the $\bm{z}$-axis, then by and angle $\pi$ about the $\bm{z}$-axis and finally back to the $\bm{z}$-axis. The closed path subtends a solid angle $\Omega = \pi (1- \cos \theta_{G_{n_x}})$ . 
} 
\label{Wenfig1}
\end{figure}
 
\begin{figure}
 \begin{center}
    \includegraphics[width=8.0cm]{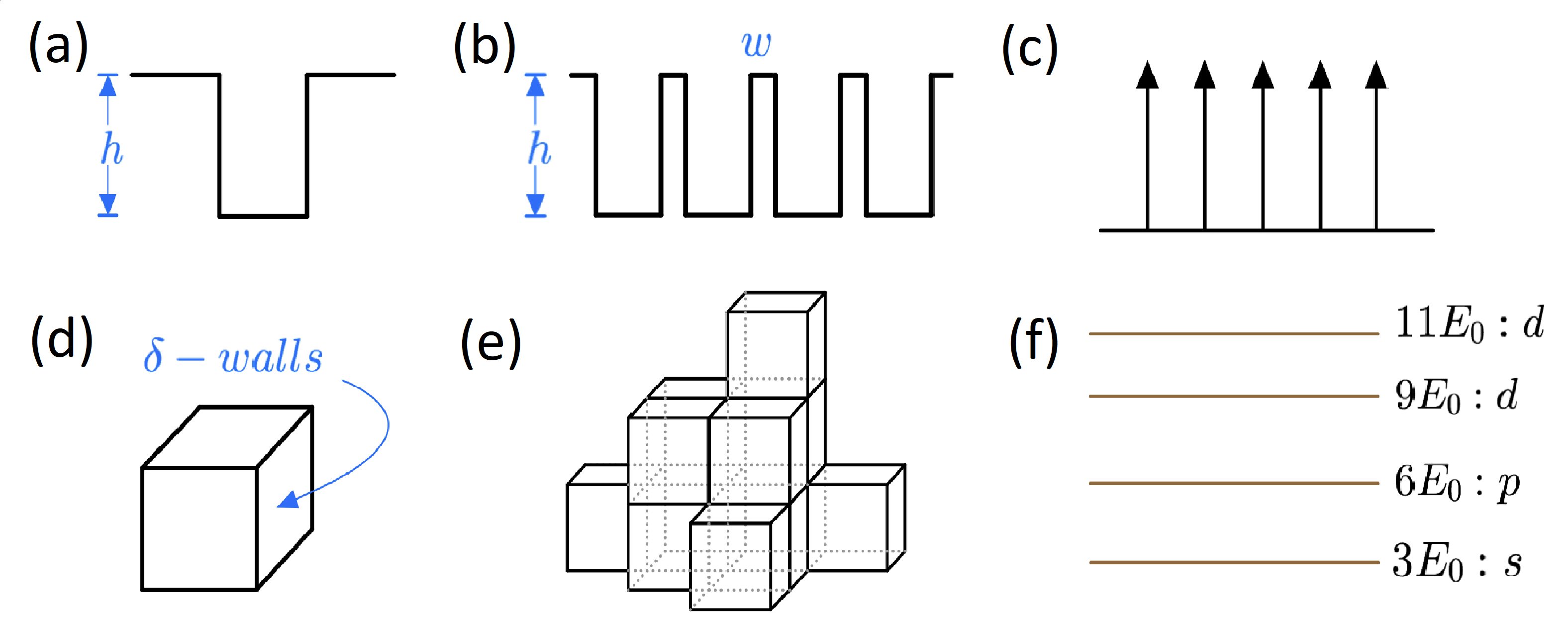}
\end{center}
  \vspace*{-5pt}  
        \caption{\small (a) An atom comprises a region of potential $V(x) =0$. The surrounding vacuum is at $V(x ) = h$. (b) Corresponding to the  interstitial region, there is a barrier height $h$ and width $w$ between adjacent the atoms. (c) As  $h \to \infty$ with the product $hw = V$ a constant the interstitials become $\delta$-functions. (d) In three dimensions the atoms are cubes separated by $\delta$-function walls and can be stacked to make a solid as shown in (e). For $V \to \infty$ the standing waves with $E_{s_x,s_y,s_z} = \frac{\hbar^2}{2m} (\frac{\pi}{a})^2 ({s_x}^2 + {s_y}^2 +{s_z}^2 ) $ have the energy scheme shown in (f). The states with two nodes are split into two non-degenerate triplets with a crystal field splitting $2E_0$.  
}  \vspace*{-10pt}  
        \label{Wenfig2}
\end{figure}

{\it Spin Berry phase physics.} As explained in any number of texts, see e.g., Ref.~\cite{Shankar1994} the original such phase\cite{Berry1984} corresponds to an adiabatic rotation of an applied magnetic field within a time dependent Hamiltonian. However as exemplified by the Stern problem\cite{Shankar1994}  a spin Berry phase can arise for a time independent Hamiltonian simply due to the electron motion within a textured effective magnetic field\cite{Barnes2012,Ye2001a,Ye2001b,Ye2001c}. 
In Fig.~\ref{Wenfig1}a, for the free electron Rashba effect, the spin direction is in-plane and perpendicular to the momentum $\vec p$. When the system is magnetic, as  in Fig.~\ref{Wenfig1}b, this direction makes a finite angle $\theta$ to the direction of a perpendicular magnetisation\cite{Barnes2014}. Since $\theta$ depends upon $\vec p$, 
even for a uniform magnetisation $\vec M = M \hat{\bf z}$, there is a spin ``texture'' in momentum space, i.e., an electron sees an internal field that is not  parallel to $\vec M$. However for a given momentum $\vec p$, the angle $\theta$  is uniform and there is no texture in real space. This is no longer the case when a periodic potential is introduced. In the usual manner, exemplified here in a single dimensional, the spin-up wave-function $\psi_{q_x\uparrow}(x) = \sum_{n_x} e^{-i\theta_{G_{n_x}}s_y/\hbar} u(q_x+G_{n_x} )  e^{i(q_x+G_{n_x} ) x} \lvert  \uparrow \rangle$ is written in terms of a crystal moment $\hbar q_x$ and the reciprocal lattice vectors $G_{n_x} $ where $n_x$ is an integer band index and $ \lvert  \uparrow \rangle$ the spin up ket. The tilt angle $\theta \to \theta_{G_{n_x}}$ is different for different $n_x$ because this is determined by $ p_x = \hbar (q_x + G_{n_x} )$ and not just $\hbar q_x $. 
Then, since the phase factor $e^{-i\theta_{G_{n_x}}s_y/\hbar}$ involves $s_x$, the electron spin direction depends upon the real space position within the unit cell. It is this that introduces a non-trivial spin Berry phase. Just how this works is taken up again below and in the Supplementary Materials\cite{SupMat}.

{\it The Kronig-Penney-Nearing model.}
The Kronig-Penney  is  a simple model\cite{Kronig1931} that connects the free electron to the tight binding model. As in Fig.~\ref{Wenfig2}a, in a one dimensions, a single atom comprises a square potential well relative to the vacuum. When many atoms are put together, as in Fig.~\ref{Wenfig2}b, each atom is separated by a barrier corresponding to the interstitial region. If this region is shrunk to zero while at the same time increasing the height of the barrier, the result is  a periodic positive delta function potential $V(x) = {V}{a} \sum_n \delta(x - na)$ illustrated in Fig.~\ref{Wenfig2}c. With decreasing $V$,  individual atomic levels broaden into tight binding bands, then describe 
 nearly free electrons for small $V$ and finally  free electrons when $V = 0$. Following  Nearing\cite{Nearing} this is generalised to higher dimensions Fig.~\ref{Wenfig2}d by assuming a separable potential, e.g., for three dimensions $V(x)+V(y) + V(z)$. Now the atoms correspond to boxes separated by delta function {\it walls\/} and can be stacked as illustrated in Fig.~\ref{Wenfig2}e. This model is separable, i.e., the 
  band energy for a crystal wave vector $\vec q$ is $E(q_x) + E(q_y) + E(q_z) $ where $E(q_x)$ is a solution to the one dimensional model. 
  
 {\it Atoms.} Much can be understood from considering three dimensional atoms, i.e., the limit $V\to \infty$. In the absence  spin-orbit coupling, the ``atomic'' energy levels $E_{n_x,n_y,n_z} = \frac{\hbar^2}{2m} (\frac{\pi}{a})^2 ({s_x}^2 + {s_y}^2 +{s_z}^2 ) $ where the integers $s_x$, $s_y$ and $s_z = 1,2,3, \ldots$. Identified is a node-less  ``s-state''  with $s_x, s_y$ and $s_z = 1$.  The energy is $3E_0$ where  $E_0 = \frac{\hbar^2}{2m}(\frac{\pi}{a})^2 $ and the band index is $n=1$. There are three degenerate $E_p = 6E_0$ single node ``p-states'' $(2,1,1)$, $(1,2,1)$ and $(1,1,2)$. The ``d-states'', with two nodes, are crystal field split into two triplets, i.e., $(1,2,2)$, $(2,1,2)$ and $(2,2,1)$  with energy 
 $E_t = 9E_0$ and  $(3,1,1)$, $(1,3,1)$ and $(1,1,3)$ with  an energy $E_g = 11E_0$. The crystal field splitting  is therefore $2 E_0$ as shown in the level scheme Fig.~\ref{Wenfig2}f. While the 1$s$ radial wave-function with no nodes can be identified with $(1,1,1)$ the higher non-degenerate n$s$-states map to the odd $(2n-1,2n-1,2n-1)$. All even combinations, e.g. $(2,2,2)$ cannot be identified as $s$-states since they have a node at the centre of the atom. It follows the 6s-orbital of Au, with five radial nodes is properly identified as $(11,11,11)$ with energy $363E_0$. Clearly the identification of e.g., higher in energy ``$p-$", ``$d-$" and ``$f-$"atomic states is an approximate concept given the atomic states here are a representation of the cubic and not rotational group.

\begin{figure}
% \vglue -10pt
  \begin{center}
    \includegraphics[width=3cm]{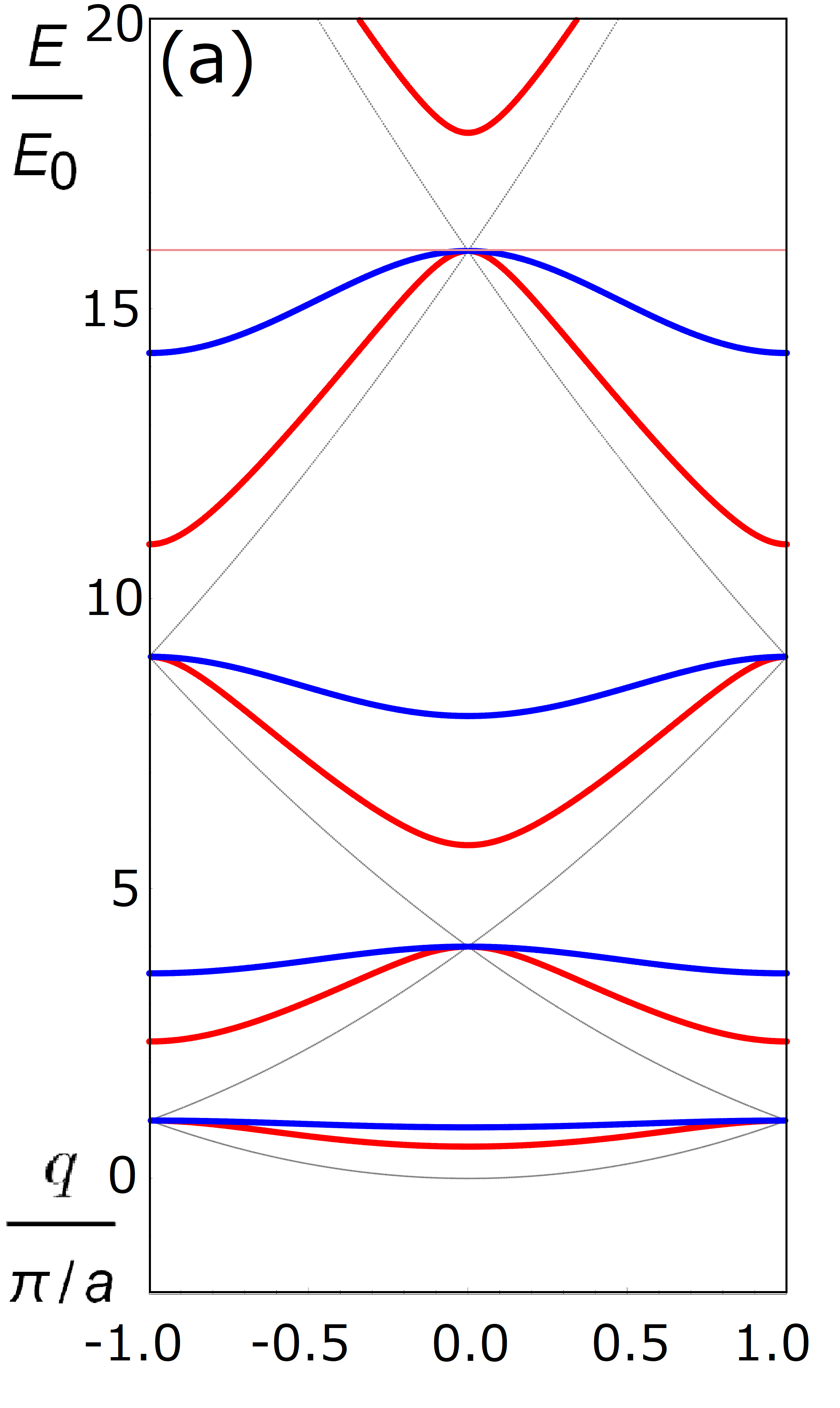}
     \includegraphics[width=4.5cm]{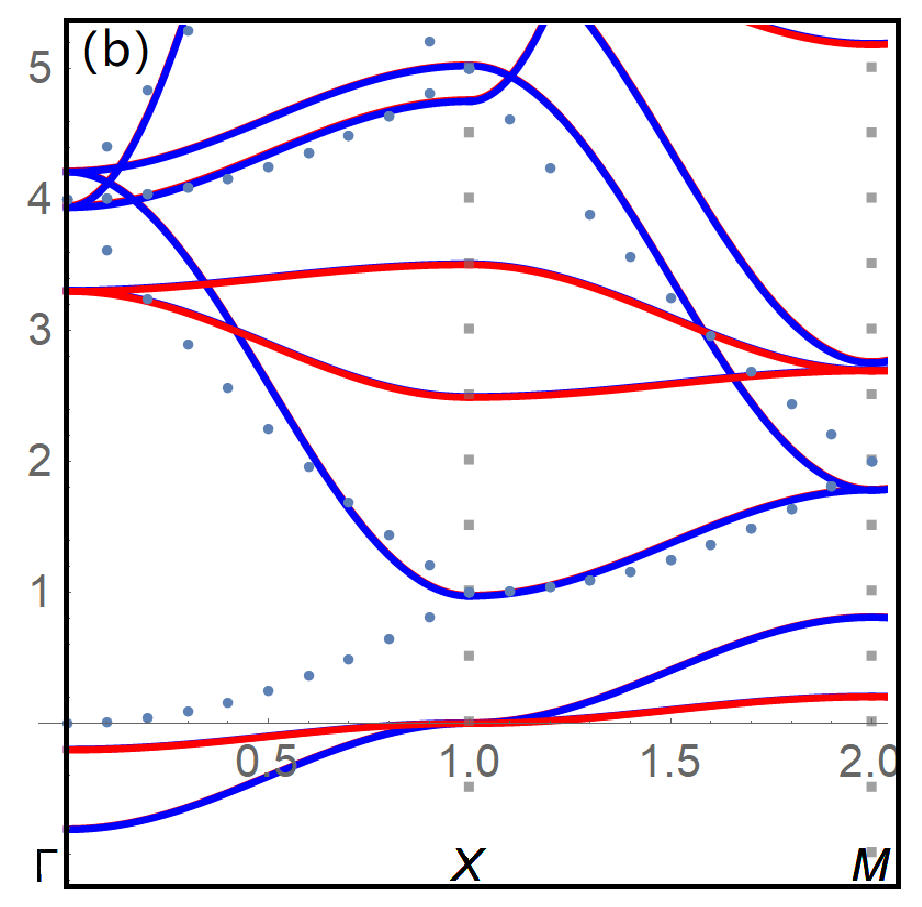}
\end{center}
        \caption{\small  (a)  Free,  nearly free ($V=E_0$),  and tight binding results  ($V=4E_0$) shown in gray, red and blue. (b) Similar results in two directions for the directions $\Gamma$-X-M in first zone. %The results are obtained using Mathematica. 
}
  
        \label{Wenfig3}
\end{figure}

{\it From free electrons to tight binding 
}\
As a touchstone for what follows, consider band formation in one and two dimensions. In Fig.~\ref{Wenfig3}a the red line correspond to nearly free electrons while the blue line agrees well with the tight binding result\cite{Tight} $E_{q_x}  = E_{n_x} - (-1)^n E_{n_x}  \frac{4E_0} {\pi^2 V}\cos q_x a $. Here, and in other figures, the fine grey lines correspond to free electrons.  In Fig.~\ref{Wenfig3}b are shown bands along high symmetry directions in two dimensions\cite{SupMat}.  %Calculation details are found in the Supplementary Materials. 
 
\begin{figure}
\begin{center}
  \includegraphics[width=7.55cm]{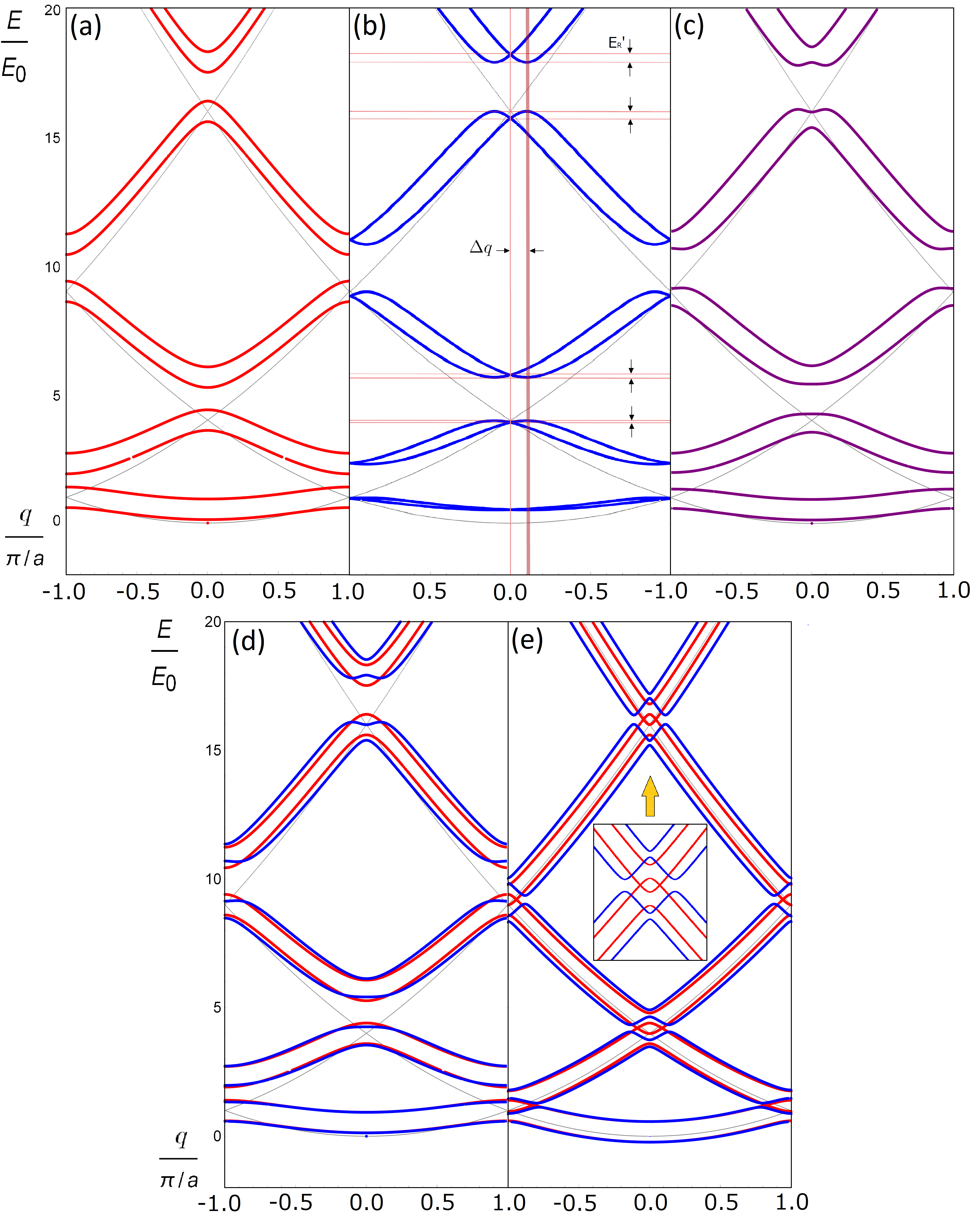}
\end{center}
     \caption{\small Lattice Rashba effect. (a) Ferromagnetic case with $V= E_0$ and $J_0S = 0.4 E_0$. The bands are shifted up and down by $\pm J_0 S$. (b) Non-magnetic Rashba results with $\alpha_R =0.2 \frac{E_0}{\pi/a} $ showing horizontal shifts by $\Delta q_x$. That this shift is independent of the band index $n_x$ is shown by the vertical lines. The horizontal lines illustrated that the empirical value of ``${E_R}$"  depends strongly on $n_x$. The actual energy shift $E_R$ is however  $n_x$ independent. (c) In plane results with $\alpha_R = 0.2\frac{E_0}{\pi/a}$  and $J_0S = 0.4E_0$. There is a spin-splitting in addition the band shifts. (d) Perpendicular ferromagnetic results. This is similar to (a) but, as can be seen by comparing the red and blue curves, the effective exchange is larger due the vector addition of the Rashba and exchange fields.
(e) Similar to (d) but with a smaller crystal potential $V = 0.2 E_{0}$. There are a double gap structures. The first order in $E$ gaps occur away from the high symmetry points ($q =0$ and $q=\pm \frac{\pi}{a}$). The value of $E$ is doubled between the blue and red curves.
}
 % \vspace*{-15pt}  
        \label{Wenfig4}
\end{figure}

{\it Stoner magnetism.
}
The problem remains fully separable when the simplest exchange potential $-J_0\vec \sigma \cdot \vec S$ is added\cite{Stoner1948}. Here,  $\vec S$ is the effective spin, per atom, of the order parameter.  As shown in Fig.~\ref{Wenfig4}a  the minority/majority bands undergo  rigid shifts  $\pm J_0S$ relative to those of Fig.~\ref{Wenfig3}a.

{\it  Spin-orbit coupling
}
As already emphasised, 
because the atomic potential is flat the full spin-orbit coupling $V_{\rm so}$ simplifies to a Rashba term $\frac{\alpha_R}{\hbar} (\vec \sigma \times \vec p)\cdot {\hat {z}}$  in which importantly $\vec p$ is the canonical momentum and where $\alpha_R = - \frac{ \hbar \lambda_c}{8\pi mc} E $ is proportional to the ``applied" and/or intrinsic  $E$ fields.

{\it  Essential physics in one dimension.
}\
The physics behind a PMA, including the effects of the Berry connection $\vec A_s$, can   be 
clearly demonstrated in one dimension. As shown in Fig.~\ref{Wenfig1}d the thin film is considered to be a  series of  chains parallel to the $x$-direction,  with an applied electric field $\vec E = E\hat{\bf z}$ in the perpendicular direction. The ``perpendicular" case has $\vec M = M\hat{\bf z }$ parallel to $\vec E$ leading to a 
Rashba field $\vec B_R$  in the $y$-direction. For the strictly one dimensional model this is degenerate with 
$\vec M =  M\hat{\bf x }$. 
This degeneracy is  lifted by adding chains in the $y$-direction as in Fig.~\ref{Wenfig1}c. If the density of electrons on these extra chains is lower, than $\vec M = M\hat{\bf x }$ corresponds to an in-plane easy axis. This two chain model is clearly separable, however, the true two dimensional problem is only approximately separable for small $\alpha_R$\cite{SupMat}. 

{\it Non-magnetic Rashba effect.} The Rashba field $\vec B_R = (\alpha_R p_x/\hbar )\hat{\bf y}$ is perpendicular to  $\vec E = E \hat{\bf z}$ putting the axis of quantisation along the $\hat {\bf x}$-axis. As indicated in Fig.~\ref{Wenfig4}b, the  bands are shifted to the left or right by $\Delta q= m \alpha_R/\hbar^2$, linear in $E$. That this shift is independent of the band index $n_x$ is illustrated by a vertical line. Also independent of $n_x$, the energy is lowered by $E_R = \frac{\hbar^2}{2m} {\Delta q}^2= \frac{1}{2} m { \alpha_R}^2/\hbar^2$. For free electrons  $E_R$ can be  determined graphically  as the vertical distance between the horizontal lines as indicated. This graphical effective ``${E_R}$" defined  in \ref{Wenfig4}b, is  correct only for the lowest $n_x =1$ band, and is almost indiscernible in Fig.~\ref{Wenfig4}b. The much larger ``${E_R}$" reflects the changing band masses. That even the highest band is lowered by only the free electron value $E_R =\frac{1}{2} m { \alpha_R}^2/\hbar^2$ is verified by a common horizontal line in Figs.~\ref{Wenfig4}b and Fig.~\ref{Wenfig3}a, i.e., with and without the Rashba effect.  It is implied the experimental $E_R \approx 3.5$meV for Au {\it cannot\/} be identified with $\frac{1}{2} m { \alpha_R}^2/\hbar^2$ and hence used to determine $\alpha_R$. The real value is much smaller.

{\it In-plane magnetisation\/}.  Take  $\vec M = M\hat{\bf x }$ so that both the exchange $\vec B_E = \pm JS_0\hat{\bf y} $ and Rashba $\vec B_R = (\alpha_R p_x/\hbar )\hat{\bf y}$ fields are along the $y$-direction and add  to give an 
apparent exchange  $JS = J_0S \pm \alpha_R p_x$.   However, e.g., for a majority electron aligned along $\vec M$, the energy $\alpha_R p_x$ can be absorbed into $\frac{\hbar^2}{2m} (p_x - m \alpha_R/\hbar^2)^2 -E_R$ where, as for the non-magnetic case,  there is,  Fig.~\ref{Wenfig4}c, a horizontal shift by $\pm \Delta q_x $ and a band index  $n_x$ independent downward energy shift $E_R$, this in addition to the exchange splitting $\mp J_0 S$. Again the evident dependence on the band index $n_x$ arises from the changes in the effective mass. As before, the energy gain is $E_R$ per electron and not the apparent ``${E_R}$". The energy shifts are second order in the applied field $\vec E$.

A {\it perpendicular magnetisation.} It is important to distinguish between the canonical $\vec p$ and crystal momentum $\hbar \vec q$. Here
$\vec M = M\hat{\bf z}$ is parallel to $\vec E$. The Rashba field $\vec B_R = (\alpha_R p_x/\hbar )\hat{\bf y}$ is perpendicular to both $\vec M$ and $\vec E$. As indicated in the lower part of Fig.~\ref{Wenfig1}c, this field adds vectorially to the exchange field $\vec B_E = J_0 S \hat{\bf z}$. The result is a larger effective exchange $JS =\sqrt{(J_0S)^2 + (\alpha_R /\hbar)^2{p_x}^2}$ for an electron of momentum $ p_x$. 
The  axis of quantisation makes an angle $\theta_{p_x} = \tan^{-1} [(\alpha_R p_x/\hbar)/(J_0S)]$ to the direction of $\vec M$. In Fig.~\ref{Wenfig4}d, in red, is reproduced Fig.~\ref{Wenfig4}a while in blue is the result including the Rashba effect. Except near the points where the free electron bands cross, the 
results are well reproduced by the analytic tight binding solution\cite{SupMat}, $ E^\pm_{n_x}(q) = E_{n_x}-  \frac{4E_0E_{n_x} } {V}\cos qa  \mp \left(  J_0 S +  {\alpha_{R}}^{2} \frac{\hbar^4 }{8J_0 S}   {G_{n_x}}^2 \right)$. The  effective exchange splitting for  the blue curve is indeed manifestly larger than that of the red curves.  

The Rashba effect increases  dramatically for higher energy bands reflecting   the increase in the reciprocal lattice vectors $G_{n_x}$ for larger band index $n_x$. This is one of the principal results and is a consequence of that fact that  the Rashba effect for the full $V_{\rm so}$ involves the canonical  $\hbar p_x$ and not just the  crystal   momentum $\hbar q_x$. 

For the surface state of Au $q_F \approx \frac{1}{10}\frac{2\pi}{a}\approx 1.5$\AA. The Rashba field is  $B_F = \alpha_R q_F \approx  \alpha_R\frac{1}{10}\frac{2\pi}{a}$ for the $n_x=1$  ``s-band''. In contrast the $n_x=3$ ``d-band'', with the same crystal momentum, has a Rashba field proportional to 
$\frac{2\pi}{10a} + 3 \frac{2\pi}{a} \approx  3 \frac{2\pi}{a}$. The splitting in  Fig.~\ref{Wenfig4}d is proportional to ${B_R}^2$ and almost $10^3$ times larger. This should be compared with the discussion of Fig.~\ref{Wenfig4}c corresponding to an in-plane magnetisation. In that case the energy gain was $E_R$ per electron independent of the band index $n_x$. As a consequence  for magnetic thin films the Rashba spin-orbit splitting leads to unambiguously to a PMA. The near cancellation for two dimensions noted in \cite{Barnes2014} no longer occurs. The Rashba field in Tesla is estimated using  $B_R = \frac{v}{2c} \frac{E}{c}$. The Au-Au separation is $\approx 2.84$\AA, but  the interstitial region that has been shrunk to zero and the model does not localise the electrons as strongly as does a central potential. Using half this distance gives $v \approx \frac{c}{5}$ and  $B_R\approx \frac{E}{10c}$. For Au the experimental $B_R \approx 5 \times 10^2$T corresponding to an internal field of $E \sim 10^{12}$V/m. The break down field of a good gate oxide, e.g, Al$_2$O$_3$ is $E\sim 10^9$V/m and sets the scale, although the local electrical field can easily be $10^{10}$V/m corresponding to $B_R \sim 3$T. In order the realise a PMA field of this magnitude needed is a small proximity exchange field such that $B_E= J_0S\approx B_R$.

{\it Electrically controllable  gaps.} The effects of the Berry phase physics are evident in Fig.~\ref{Wenfig4}d. The tight binding result is clearly inadequate in the region near the crossing of the free electron bands. There is no longer a simple increase in the effective exchange and the results (blue) with the Rashba effect cross those (red) for which this is absent. The origin of this effect is evident from the insert in Fig.~\ref{Wenfig4}e. In the absence of the Rashba coupling, there are crossings, either side of $q_x =0$, of  red pair of curves with the upper and lower lines corresponding to majority/minority bands. Reflecting the Berry phase factor $e^{-i\theta_{G_{n_x}}s_x/\hbar}$ in $\psi_{q_x\uparrow}(x) $ there are matrix elements proportional to $\sin \theta_{G_{n_x}} \sim \frac{\alpha_RG_{n_x}}{J_0S}$ that connect the majority/minority bands when an electron scatters  off the periodic potential $V(x)$. This produces gaps\cite{SupMat} that go as $V\theta_{G_{n_x}} \sim V  \frac{\alpha_RG_{n_x}}{J_0S}$ and since $\alpha_R$ is proportional to the electric field $E$, so are these gaps. % Details of calculations that support these conclusions are found in the Supplementary Materials. 
While in Fig.~\ref{Wenfig4}e there are two distinct gaps at the majority/minority crossing, with an increasing periodic potential $V $ these gaps broaden and merge as in Fig.~\ref{Wenfig4}d. Reflecting the dependence on $G_{n_x}$, there is a very strong increase in these gaps with increasing energy and band index $n_x$. In Figs.~\ref{Wenfig4}d and Fig.~\ref{Wenfig4}e, the Berry phase gap goes from being barely visible for $E \sim 2E_0$ and $q_x \lesssim \frac{\pi}{a}$, to being comparable to that due to the periodic potential at $E \sim 4E_0$ and $q_x \approx 0$, and larger than the periodic gap for $E \sim 9E_0$ and $q_x \lesssim \frac{\pi}{a}$. That for  higher energies the gap hardly increases, e.g., from $E \sim 9E_0$ to $16E_0$  reflects the fact that 
 the Rashba energy $\alpha_RG_{n_x= 3}$ is comparable to  $J_0S$, the exchange splitting. 
Again these changes are due to the fact that the Rashba effect reflects the full momentum $p_x =\hbar( q_x + G_{n_x})$ and not just the crystal momentum $\hbar q_x$. For a magnetic proximity effect  engineered to have $J_0 S\approx \alpha_RG_{n_x}$ ($B_E\approx B_R$) the effective field is $\sim V$ in energy units and easily of the order of electron volts. In the illustrative calculations, of this and previous sections, the value of $\alpha_R$ is unrealistically large and chosen to make evident the various effects.

{\it Peierls effect.} These electrically controlled gaps occur at critical values of the wave-vector $\vec q = \vec q_c$ at which occurs an accidental crossing of minority/majority bands. That such gaps lie at the Fermi energy $E_F$ and produce a PMA linear in $E$ is therefore very much material dependent. However a Peierls effect can occur in which $\vec M$ tilts an angle $\alpha$ away from the vertical thereby  creating a shift $\Delta q$, proportional to the in-plane component of $\vec M$, that moves $\vec q_c$ to the Fermi surface. In this manner there is not only an electrical control of the PMA but also of the tilt angle. Since the spin torque transfer (STT) effect\cite{Slonczewski1996,Barnes2012} is absent, and highly dissipative, for a strictly perpendicular $\vec M$, this is of considerable technological importance.

% You, Long and Lee, OukJae and Bhowmik, Debanjan and Labanowski, Dominic and Hong, Jeongmin and Bokor, Jeffrey and Salahuddin, Sayeef
% Switching of perpendicularly polarized nanomagnets with spin orbit torque without an external magnetic field by engineering a tilted anisotropy
% Proceedings of the National Academy of Sciences {bf 112} 10310-10315

%R. Karplus and J. M. Luttinger, Phys. Rev. 95, 1154 (1954).

%Yugui Yao1,2,3, L. Kleinman1, A. H. MacDonald1, Jairo Sinova4,1, T. Jungwirth5,1, Ding-sheng Wang3, Enge Wang2,3, Qian Niu1 First Principles Calculation of Anomalous Hall Conductivity in Ferromagnetic bcc Fe

%This is distinct from the Berry phase effect in real space discussed in J. Ye, et al., Phys. Rev. Lett. 83, 3737 (1999); Y. Taguchi, et al., Science 291, 2573 (2001); R. Shindou, et al., Phys. Rev. Lett. 87, 116801 (2001), where the AHE is related with non collinear spin-splitting effective fields produced by spin-lattice backgrounds.

{\it Mathematical methods.\/} Schr\"odinger's equation reduces to\cite{SupMat}: 
\begin{equation}
 \frac{1}{V} =\sum_{n_x} {\cal G}_{n_x} (E_{q_x} )
 \label{E101}
\end{equation}
where $V$ is the strength of the delta function potentials. This is
written in terms of the free electron Green's function, e.g.,  with a  perpendicular magnetisation ${\cal G}_n (E_{q_x} ) = e^{-i\theta_{G_{n_x}}s_x/\hbar}
(E_{q_x} -\frac{\hbar^2(q_x+G_{n_x})^2}{2m} - \sigma_z {JS}
)^{-1}e^{i\theta_{G_{n_x}}s_x/\hbar}$, a two-by-two matrix. Exploiting  the SU(2) algebra, Eq.~\ref{E101} reduces to the form $\frac{1}{V} = {\cal S} +  {\cal A}  \sigma_{z}  + {\cal B}  \sigma_{y}$ and finally to an implicit secular equation for $E_{q_x} $: $
        \left(\frac{1}{V}-{\cal S}\right)^2 
        - \left({\cal A}^2
        +{\cal B}^2\right)=0
$    
with rather complicated definitions of the three functions ${\cal S}$,  ${\cal A}$ and ${\cal B}$, see\cite{SupMat}.  This non-linear equation has many solutions for a given $q_x$. These are found using Mathematica and are assigned a band index $n_x$ by comparison with the atomic limit. 

{\it Approximate separation of variables in two  dimensions.
} 
Since the spin orbit terms $\frac{\alpha_R }{\hbar}  \sigma_x p_y$ and $ - \frac{\alpha_R }{\hbar} \sigma_y \vec p_x$ act on the same spin ket $\lvert \rangle$ an exact separation is not possible. This notwithstanding\cite{SupMat} if the Rashba field $B_R$ is small compared with exchange field $B_E$, defined is separation {\it matrix}, e.g., $ M_x= \frac{1}{\psi_x (x) }\hat H_x \psi_x (x) $ containing three independent constants obtained by requiring ${\cal S}$,  ${\cal A}$ and ${\cal B}$ all to be zero. The matrix $ M_y$ is similarly defined  and finally  $[(M_x + M_y) - E] \lvert \rangle =0$, where $E$ is the energy, is solved for the spin ket $\lvert \rangle $. Except for the gaps, the essentially physics is contained in  the two dimensional tight binding results. 
The net PMA energy $
 {\alpha_{R}}^{2} \frac{\hbar^4 }{8J_0 S}   {G_{n_y}}^2  - \frac{\hbar^2}{2m} \frac{m^2}{\hbar^4}{\alpha_R}^2 
$
is dominated by the term proportional to the $ {G_{n_y}}^2 $. There is a ``enormous'' $E^2$ perpendicular anisotropy energy PMA as reported e.g., by Lau et al \cite{Lau2019}.

{\it A different  Berry phase.} The introduction of a Berry phase has become quite popular. The Berry phase physics introduced here is different. In particular the Berry connection $\vec A_s$, i.e., an effective vector potential defined elsewhere\cite{Barnes2012} contains an off-diagonal part  that is invariably ignored. It is this part of $\vec A_s$ that leads to the electrically controlled gaps described above. This is taken up again immediately below.

 The spin-orbit coupling implies a Luttinger anomalous velocity\cite{Luttinger1954} and this leads to  Berry phase  corrections to  the dynamical equations of motion.  As described by Yao et al.\cite{Yao2004} these dynamics can be scotched onto a relativistic band calculation. Invariably the band structure package, in this case WIEN2K\cite{Blaha2001}, includes only the effects of a central potential. In another approach\cite{Ye2001a,Ye2001b,Ye2001c}  a spin Berry phase reflects a spin texture, i.e., non-co-linear magnetisation $\vec M(\vec r;t)$. In the present development the central potential $V(r)$ is absent and the order parameter  $\vec M(\vec r;t)$ is uniform. Again, this new Berry phase physics  cannot be captured in the current flavours of the DFT\cite{Kresse1993a,Kresse1993b,Kresse1993c,Kresse1993d,MacDonald2018}.

{\it Berry phase physics - perpendicular magnetisation.}   In this case, Eq.~\ref{E101} becomes
\begin{equation}
\frac{1}{V} =
           % \frac{1}{\sqrt{2}} 
            \sum_{n_x,\pm}
              (1  \pm   e^{ -i \frac{\pi}{2} (\cos \theta_{G_{n_x}} +\sigma_z)\sigma_z + \sin \theta_{G_{n_x}} \sigma_y) } )
           % e^{\pm i\frac{\pi}{2}(\cos \theta_{G_{n_x}} \sigma_z + \sin \theta_{G_{n_x}} \sigma_y)}  
            {\cal G}^{\pm}_{G_{n_x}}
            \end{equation}
 where $ {\cal G}^{\pm}_{n_x} = [E_{q_x} -\frac{\hbar^2(q_x+G_{n_x})^2}{2m}\pm {JS}]^{-1}$ exhibits the exchange splitting $JS$ including the Rashba correction\cite{SupMat} and corresponds to the dynamic phase. The phase factor $ 
 e^{ -i \frac{\pi}{2} (\cos \theta_{G_{n_x}} +\sigma_z)\sigma_z + \sin \theta_{G_{n_x}} \sigma_y) } 
   %e^{\pm i\frac{\pi}{2}(\cos \theta_{G_{n_x}} \sigma_z + \sin \theta_{G_{n_x}} \sigma_y)} 
   $ is that generated\cite{SupMat} by the Berry connection\cite{Barnes2012}  $\vec A_s$ for the spin path shown in Fig.~\ref{Wenfig1}g that involves a $\phi$ rotation of $\pi$ at an angle $ \theta_{G_{n_x}}$. Consider a 
  more
  intuitive two dimensional example. As shown in Fig.~\ref{Wenfig1}e, a real space path of an electron maps to one on the Bloch sphere. In Fig.~\ref{Wenfig1}f, an electron executes a closed path by undergoing elastic reflections with four crystal planes. The spin always makes the same  angle $\theta$ to the direction of $\vec E = E\hat{\bf z}$ but, since the momentum has a different direction, the angle $\phi$ changes by $\frac{\pi}{2}$ at each reflection. There is a solid angle $\Omega $ subtended  by the path on the Bloch sphere leading to a diagonal Berry phase $\frac{1}{2} \Omega $.

{\it Advancing to the Iron Age.} The Lieb proof\cite{Lieb1983} of the  Hohenberg-Kohn theorem\cite{Hohenberg1964} and the resulting Kohn-Sham equations\cite{Kohn1965} do not admit a non-trivial Berry connection $\vec A_s$. The constrained diamagnetic DFT\cite{Kohn2004} {\it does\/} admit an electromagnetic Berry connection $\vec A$ as an {\it extrinsic\/} potential\cite{Vignale1988} and hence admits a magnetic field. For a  fraction $\frac{p}{q}$ of the flux quantum per unit cell, a Hofstadter butterfly\cite{Janecek2013} occurs in which the 
 unit cell is $q$ times larger. Even this possibility is not included in current flavours of the DFT. Worse, the spin Berry connection $\vec A_s$ introduced here\cite{Barnes2012} is {\it intrinsic\/} whence this extension is not valid. 
In Greek mythology the Golden Age  leads  finally to the Iron Age. It might be hoped the much heralded  materials golden age based on the  DFT\cite{Adler2019} will  soon describe the essential Berry phase physics of capped iron  thin-films with a Rashba PMA. 

{\it  Summary.}  When compared with the free electron model\cite{Barnes2014}, accounting for the periodic potential leads an enormous Rashba PMA proportional to $E^2$. Illustrating  the importance of the usually ignored off-diagonal part of the  spin Berry connection\cite{Barnes2012} $\vec A_s$, found are  electrically controlled band structure  gaps proportional to $E$ with the promise of a truly giant, linear in $E$, electrical control of magnetism accompanied by a spin Peierls effect that implies low dissipation dynamics.


\begin{thebibliography}{10}
\providecommand{\url}[1]{{#1}}
\providecommand{\urlprefix}{URL }
\providecommand{\doi}[1]{\url{https://doi.org/#1}}
\bibcommenthead

\bibitem{Wang2020}
L.~Wang, X.~Li, T.~Sasaki, K.~Wong, G.~Yu, S.~Peng, C.~Zhao, T.~Ohkubo,
  K.~Hono, W.~Zhao, K.~Wang, High voltage-controlled magnetic anisotropy and
  interface magnetoelectric effect in sputtered multilayers annealed at high
  temperatures.
\newblock Science China Physics, Mechanics {\&} Astronomy \textbf{63}(7),
  277,512 (2020).
\newblock \doi{10.1007/s11433-019-1524-y}.
\newblock \urlprefix\url{https://doi.org/10.1007/s11433-019-1524-y}

\bibitem{Barnes2014}
S.E. Barnes, J.~Ieda, S.~Maekawa, Rashba spin-orbit anisotropy and the electric
  field control of magnetism.
\newblock Scientific Reports \textbf{4}(1), 4105 (2014).
\newblock \doi{10.1038/srep04105}.
\newblock \urlprefix\url{https://doi.org/10.1038/srep04105}

\bibitem{Barnes2012}
S.E. Barnes, in \emph{Spin Current}, ed. by S.~Maekawa, S.O. Valenzuela,
  E.~Saitoh, T.~Kimura (Oxford Scholarship, 2013), chap.~7

\bibitem{Gruner1988}
G.~Gr\"uner, The dynamics of charge-density waves.
\newblock Rev. Mod. Phys. \textbf{60}, 1129--1181 (1988).
\newblock \doi{10.1103/RevModPhys.60.1129}.
\newblock \urlprefix\url{https://link.aps.org/doi/10.1103/RevModPhys.60.1129}

\bibitem{Barnes1983}
S.E. Barnes, A.~Zawadowski, Theory of josephson-type oscillations in a moving
  charge-density wave.
\newblock Phys. Rev. Lett. \textbf{51}, 1003--1006 (1983).
\newblock \doi{10.1103/PhysRevLett.51.1003}.
\newblock \urlprefix\url{https://link.aps.org/doi/10.1103/PhysRevLett.51.1003}

\bibitem{Slonczewski1996}
J.~Slonczewski, Current-driven excitation of magnetic multilayers.
\newblock Journal of Magnetism and Magnetic Materials \textbf{159}(1), L1--L7
  (1996).
\newblock \doi{https://doi.org/10.1016/0304-8853(96)00062-5}.
\newblock
  \urlprefix\url{https://www.sciencedirect.com/science/article/pii/0304885396000625}

\bibitem{Messiah1966}
A.~Messiah, \emph{Quantum Mechanics} (Dover Publications, 2014)

\bibitem{FW1950}
L.L. Foldy, S.A. Wouthuysen, On the dirac theory of spin 1/2 particles and its
  non-relativistic limit.
\newblock Phys. Rev. \textbf{78}, 29--36 (1950).
\newblock \doi{10.1103/PhysRev.78.29}.
\newblock \urlprefix\url{https://link.aps.org/doi/10.1103/PhysRev.78.29}

\bibitem{Winkler2003}
R.~Winkler, {Spin-orbit coupling effects in two-dimensional electron and
  Hole Systems} (Springer, 2011)

\bibitem{Wang1996}
X.~Wang, R.~Wu, D.S. Wang, A.J. Freeman, Torque method for the theoretical
  determination of magnetocrystalline anisotropy.
\newblock Phys. Rev. B \textbf{54}, 61--64 (1996).
\newblock \doi{10.1103/PhysRevB.54.61}.
\newblock \urlprefix\url{https://link.aps.org/doi/10.1103/PhysRevB.54.61}

\bibitem{Rashba1984}
Y.A. {Bychkov}, {\'E}.I. {Rashba}, {Properties of a 2D electron gas with lifted
  spectral degeneracy}.
\newblock Soviet Journal of Experimental and Theoretical Physics Letters
  \textbf{39}, 78 (1984)

\bibitem{Ong2015}
P.V. Ong, N.~Kioussis, D.~Odkhuu, P.~Khalili~Amiri, K.L. Wang, G.P. Carman,
  Giant voltage modulation of magnetic anisotropy in strained heavy
  metal/magnet/insulator heterostructures.
\newblock Phys. Rev. B \textbf{92}, 020,407 (2015).
\newblock \doi{10.1103/PhysRevB.92.020407}.
\newblock \urlprefix\url{https://link.aps.org/doi/10.1103/PhysRevB.92.020407}

\bibitem{Yao2004}
Y.~Yao, L.~Kleinman, A.H. MacDonald, J.~Sinova, T.~Jungwirth, D.S. Wang,
  E.~Wang, Q.~Niu, First principles calculation of anomalous hall conductivity
  in ferromagnetic bcc Fe.
\newblock Phys. Rev. Lett. \textbf{92}, 037,204 (2004).
\newblock \doi{10.1103/PhysRevLett.92.037204}.
\newblock
  \urlprefix\url{https://link.aps.org/doi/10.1103/PhysRevLett.92.037204}

\bibitem{Manchon2015}
A.~Manchon, H.C. Koo, J.~Nitta, S.M. Frolov, R.A. Duine, New perspectives for
  Rashba spin--orbit coupling.
\newblock Nature Materials \textbf{14}(9), 871--882 (2015).
\newblock \doi{10.1038/nmat4360}.
\newblock \urlprefix\url{https://doi.org/10.1038/nmat4360}

\bibitem{Blaha2001}
P.~Blaha, K.~Schwarz, G.~Madsen, D.~Kvasnicka, J.~Luitz, R.~Laskowsk, F.~Tran,
  L.~Marks, L.~Marks, \emph{WIEN2k: An Augmented Plane Wave Plus Local Orbitals
  Program for Calculating Crystal Properties} (Techn. Universitat, 2019)

\bibitem{Kresse1993a}
D.~Hobbs, G.~Kresse, J.~Hafner, Fully unconstrained noncollinear magnetism
  within the projector augmented-wave method.
\newblock Phys. Rev. B \textbf{62}, 11,556--11,570 (2000).
\newblock \doi{10.1103/PhysRevB.62.11556}.
\newblock \urlprefix\url{https://link.aps.org/doi/10.1103/PhysRevB.62.11556}

\bibitem{Kresse1993b}
G.~Kresse, J.~Furthm\"uller, Efficient iterative schemes for ab initio
  total-energy calculations using a plane-wave basis set.
\newblock Phys. Rev. B \textbf{54}, 11,169--11,186 (1996).
\newblock \doi{10.1103/PhysRevB.54.11169}.
\newblock \urlprefix\url{https://link.aps.org/doi/10.1103/PhysRevB.54.11169}

\bibitem{Kresse1993c}
G.~Kresse, J.~Furthmüller, Efficiency of ab-initio total energy calculations
  for metals and semiconductors using a plane-wave basis set.
\newblock Computational Materials Science \textbf{6}(1), 15--50 (1996).
\newblock \doi{https://doi.org/10.1016/0927-0256(96)00008-0}.
\newblock
  \urlprefix\url{https://www.sciencedirect.com/science/article/pii/0927025696000080}

\bibitem{Kresse1993d}
D.~Hobbs, G.~Kresse, J.~Hafner, Fully unconstrained noncollinear magnetism
  within the projector augmented-wave method.
\newblock Phys. Rev. B \textbf{62}, 11,556--11,570 (2000).
\newblock \doi{10.1103/PhysRevB.62.11556}.
\newblock \urlprefix\url{https://link.aps.org/doi/10.1103/PhysRevB.62.11556}

\bibitem{Koseki2019}
S.~Koseki, N.~Matsunaga, T.~Asada, M.W. Schmidt, M.S. Gordon, Spin–orbit
  coupling constants in atoms and ions of transition elements: Comparison of
  effective core potentials, model core potentials, and all-electron methods.
\newblock The Journal of Physical Chemistry A \textbf{123}(12), 2325--2339
  (2019).
\newblock \doi{10.1021/acs.jpca.8b09218}.
\newblock \urlprefix\url{https://doi.org/10.1021/acs.jpca.8b09218}.
\newblock PMID: 30817150.
\newblock
  {\href{https://arxiv.org/abs/https://doi.org/10.1021/acs.jpca.8b09218}{{https://doi.org/10.1021/acs.jpca.8b09218}}}

\bibitem{Landau1991}
L D Landau, E.M. Lifshitz \emph{Quantum mechanics: non-relativistic theory}
  (Butterworth Heinemann, 1991)

\bibitem{SupMat}
See the supplementary materials for details of the calculations.

\bibitem{radialint}
The integral $\langle \frac{1}{r^3} \rangle = \frac{Z^3}{n^3\ell(\ell +
  \frac{1}{2})(\ell+1)} \frac{1}{{a_0}^3} $, so if $\ell = 1$ then $n=2$ and
  there is an additional factor of $\frac{1}{n^3\ell(\ell +
  \frac{1}{2})(\ell+1)} = \frac{1}{24}$. For the estimation of the electric
  field $\langle \frac{1}{r^2} \rangle =\frac{2Z^2}{n^3(2\ell+1) {a_0}^2} $ is
  used

\bibitem{Shankar1994}
R.~Shankar, \emph{Principles of Quantum Mechanics} (Springer, 2014)

\bibitem{Berry1984}
M.V. {Berry}, {Quantal Phase Factors Accompanying Adiabatic Changes}.
\newblock Proceedings of the Royal Society of London Series A
  \textbf{392}(1802), 45--57 (1984).
\newblock \doi{10.1098/rspa.1984.0023}

\bibitem{Ye2001a}
J.~Ye, Y.B. Kim, A.J. Millis, B.I. Shraiman, P.~Majumdar, Z.~Te\ifmmode
  \check{s}\else \v{s}\fi{}anovi\ifmmode~\acute{c}\else \'{c}\fi{}, Berry phase
  theory of the anomalous Hall effect: Application to colossal
  magnetoresistance manganites.
\newblock Phys. Rev. Lett. \textbf{83}, 3737--3740 (1999).
\newblock \doi{10.1103/PhysRevLett.83.3737}.
\newblock \urlprefix\url{https://link.aps.org/doi/10.1103/PhysRevLett.83.3737}

\bibitem{Ye2001b}
Y.~Taguchi, Y.~Oohara, H.~Yoshizawa, N.~Nagaosa, Y.~Tokura, Spin chirality,
  Berry phase, and anomalous Hall effect in a frustrated ferromagnet.
\newblock Science \textbf{291}, 2573–2576 (2001).
\newblock \doi{10.1126/science.1058161}

\bibitem{Ye2001c}
R.~Shindou, N.~Nagaosa, Orbital ferromagnetism and anomalous Hall effect in
  antiferromagnets on the distorted fcc lattice.
\newblock Phys. Rev. Lett. \textbf{87}, 116,801 (2001).
\newblock \doi{10.1103/PhysRevLett.87.116801}.
\newblock
  \urlprefix\url{https://link.aps.org/doi/10.1103/PhysRevLett.87.116801}

\bibitem{Kronig1931}
R.D.L. Kronig, W.G. Penney, R.H. Fowler, Quantum mechanics of electrons in
  crystal lattices.
\newblock Proceedings of the Royal Society of London. Series A, Containing
  Papers of a Mathematical and Physical Character \textbf{130}(814), 499--513
  (1931).
\newblock \doi{10.1098/rspa.1931.0019}.
\newblock
  \urlprefix\url{https://royalsocietypublishing.org/doi/abs/10.1098/rspa.1931.0019}.
\newblock
  {\href{https://arxiv.org/abs/https://royalsocietypublishing.org/doi/pdf/10.1098/rspa.1931.0019}{{https://royalsocietypublishing.org/doi/pdf/10.1098/rspa.1931.0019}}}

\bibitem{Nearing}
James~Nearing.
\newblock Unpublished

\bibitem{Tight}
V.~Rosato, M.~Guillope, B.~Legrand.
\newblock Thermodynamical and structural properties of f.c.c. transition metals
  using a simple tight-binding model (1989).
\newblock \doi{10.1080/01418618908205062}.
\newblock
  \urlprefix\url{https://www.tandfonline.com/doi/abs/10.1080/01418618908205062}

\bibitem{Stoner1948}
E.C. Stoner, E.P. Wohlfarth, A mechanism of magnetic hysteresis in
  heterogeneous alloys.
\newblock Philosophical Transactions of the Royal Society of London. Series A,
  Mathematical and Physical Sciences \textbf{240}(826), 599--642 (1948).
\newblock \doi{10.1098/rsta.1948.0007}.
\newblock
  \urlprefix\url{https://royalsocietypublishing.org/doi/abs/10.1098/rsta.1948.0007}.
\newblock
  {\href{https://arxiv.org/abs/https://royalsocietypublishing.org/doi/pdf/10.1098/rsta.1948.0007}{{https://royalsocietypublishing.org/doi/pdf/10.1098/rsta.1948.0007}}}

\bibitem{Lau2019}
Y.C. Lau, Z.~Chi, T.~Taniguchi, M.~Kawaguchi, G.~Shibata, N.~Kawamura,
  M.~Suzuki, S.~Fukami, A.~Fujimori, H.~Ohno, M.~Hayashi, Giant perpendicular
  magnetic anisotropy in Ir/Co/Pt multilayers.
\newblock Phys. Rev. Materials \textbf{3}, 104,419 (2019).
\newblock \doi{10.1103/PhysRevMaterials.3.104419}.
\newblock
  \urlprefix\url{https://link.aps.org/doi/10.1103/PhysRevMaterials.3.104419}

\bibitem{Luttinger1954}
R.~Karplus, J.M. Luttinger, Hall effect in ferromagnetics.
\newblock Phys. Rev. \textbf{95}, 1154--1160 (1954).
\newblock \doi{10.1103/PhysRev.95.1154}.
\newblock \urlprefix\url{https://link.aps.org/doi/10.1103/PhysRev.95.1154}

\bibitem{MacDonald2018}
G.~Chaudhary, M.d.S. Dias, A.H. MacDonald, S.~Lounis, Anatomy of magnetic
  anisotropy induced by Rashba spin-orbit interactions.
\newblock Phys. Rev. B \textbf{98}, 134,404 (2018).
\newblock \doi{10.1103/PhysRevB.98.134404}.
\newblock \urlprefix\url{https://link.aps.org/doi/10.1103/PhysRevB.98.134404}

\bibitem{Lieb1983}
E.H. Lieb, Density functionals for coulomb systems.
\newblock International Journal of Quantum Chemistry \textbf{24}(3), 243--277
  (1983).
\newblock \doi{https://doi.org/10.1002/qua.560240302}.
\newblock
  \urlprefix\url{https://onlinelibrary.wiley.com/doi/abs/10.1002/qua.560240302}.
\newblock
  {\href{https://arxiv.org/abs/https://onlinelibrary.wiley.com/doi/pdf/10.1002/qua.560240302}{{https://onlinelibrary.wiley.com/doi/pdf/10.1002/qua.560240302}}}

\bibitem{Hohenberg1964}
P.~Hohenberg, W.~Kohn, Inhomogeneous electron gas.
\newblock Phys. Rev. \textbf{136}, B864--B871 (1964).
\newblock \doi{10.1103/PhysRev.136.B864}.
\newblock \urlprefix\url{https://link.aps.org/doi/10.1103/PhysRev.136.B864}

\bibitem{Kohn1965}
W.~Kohn, L.J. Sham, Self-consistent equations including exchange and
  correlation effects.
\newblock Phys. Rev. \textbf{140}, A1133--A1138 (1965).
\newblock \doi{10.1103/PhysRev.140.A1133}.
\newblock \urlprefix\url{https://link.aps.org/doi/10.1103/PhysRev.140.A1133}

\bibitem{Kohn2004}
W.~Kohn, A.~Savin, C.A. Ullrich, Hohenberg–Kohn theory including spin
  magnetism and magnetic fields.
\newblock International Journal of Quantum Chemistry \textbf{101}(5), 20--21
  (2005).
\newblock \doi{https://doi.org/10.1002/qua.20303}.
\newblock
  \urlprefix\url{https://onlinelibrary.wiley.com/doi/abs/10.1002/qua.20303}.
\newblock
  {\href{https://arxiv.org/abs/https://onlinelibrary.wiley.com/doi/pdf/10.1002/qua.20303}{{https://onlinelibrary.wiley.com/doi/pdf/10.1002/qua.20303}}}

\bibitem{Vignale1988}
G.~Vignale, M.~Rasolt, Current- and spin-density-functional theory for
  inhomogeneous electronic systems in strong magnetic fields.
\newblock Phys. Rev. B \textbf{37}, 10,685--10,696 (1988).
\newblock \doi{10.1103/PhysRevB.37.10685}.
\newblock \urlprefix\url{https://link.aps.org/doi/10.1103/PhysRevB.37.10685}

\bibitem{Janecek2013}
S.~Janecek, M.~Aichinger, E.R. Hern\'andez, Two-dimensional Bloch electrons in
  perpendicular magnetic fields: An exact calculation of the hofstadter
  butterfly spectrum.
\newblock Phys. Rev. B \textbf{87}, 235,429 (2013).
\newblock \doi{10.1103/PhysRevB.87.235429}.
\newblock \urlprefix\url{https://link.aps.org/doi/10.1103/PhysRevB.87.235429}

\bibitem{Adler2019}
R.~Adler, C.J. Kang, C.H. Yee, G.~Kotliar, Correlated materials design:
  prospects and challenges.
\newblock Reports on Progress in Physics \textbf{82}(1), 012,504 (2018).
\newblock \doi{10.1088/1361-6633/aadca4}.
\newblock \urlprefix\url{https://doi.org/10.1088%2F1361-6633%2Faadca4}

\end{thebibliography}
\end{document}

\title{Supplementary Materials for \\
Giant Rashba electrical control of magnetism in band models
}

\author{ Wen Li and Stewart E.  Barnes}

\affiliation{
Physics Department, University of Miami, Coral Gables, FL 33124, USA. 
}

\hfill

%\date{\today}

%L. Z. Wang, X. Li, T. Sasaki, K. Wong, G. Q. Yu, S. Z. Peng, C. Zhao, T. Ohkubo, K. Hono, W. S. Zhao, and K. L. Wang, High voltage-controlled magnetic anisotropy and interface magnetoelectric effect in sputtered multilayers annealed at high temperatures, Sci. China-Phys. Mech. Astron. 63, 277512 (2020), https://doi.org/10.1007/s11433-019-1524-y

%Veit, M.J., Arras, R., Ramshaw, B.J. et al. Nonzero Berry phase in quantum oscillations from giant Rashba-type spin splitting in LaTiO3/SrTiO3 heterostructures. Nat Commun 9, 1458 (2018). https://doi.org/10.1038/s41467-018-04014-0

\maketitle

{\it  The Kronig-Penney model.} This is unique in that permits an analytic study  of the transition from the  free electron to the tight bonding model in one, two or  higher dimensions. The separable generalisation to higher dimensions  is due to Nearing\cite{Nearing}. 
A {\it periodic\/} delta-function corresponds to $V(x) =Va  \sum_n \delta(x-a n) =  V \sum_n e^{iG_nx}$ where the reciprocal lattice vectors are $G_n = \frac{2\pi}{a}n$, $a$ is the lattice spacing and $V$ the strength of the potential in energy units. In three   dimensions, the separable potential  $V(x)+ V(y) + V(z)$, defines boxes of side $a$, that reflect atoms separated by  interstitial walls. See Fig.~[1] of the main text.

For simplicity, and familiarity, first the  results for this model are re-derived in one dimension and generalised to the relevant two dimensions as is appropriate.  The Bloch theorem dictates the wave function is of the form $\psi_{q}(x) = e^{iq_x x} u_q(x) $ where $u_q(x) $ is a periodic function, i.e. $u_{q_x} (x) = u_{q_x} (x\pm na) $. 
 Without spin-orbit coupling the Kronig-Penney model $u_{q_x} (x) $ can be taken as real. In the case when the Rashba coupling is included it is still the case that $u_{q_x}(x) = \sum_n u_{q_x}(G_n) e^{iG_nx}$ but where $u_{q_x} (G_n) \ne u_{q_x} (-G_n) $. With periodic boundary conditions on an interval $L$, the normalisation $\frac{1}{L} \int dx| \psi_{q_x}(x)|^2 = 1$ implies $\sum_n |u_q(G_n) |^2 = 1$. For the Hamiltonian  $H = \frac{1}{2m}p^2 + V(x)$ the Schr\"odinger equation is $H  \psi_{q_x} (x) = E_{q_x}  \psi_q (x) $ or
\begin{equation}
    \left[
 - \frac{\hbar^2}{2m} \frac{d^2}{dx^2} + V a \sum_n \delta(x-na) 
 \right]
 \psi_{q_x} (x) = E_{q_x}  \psi_q (x) 
 \tag{S1}\label{S1}
\end{equation}
Corresponding to Bloch's theorem, for any $q_x$, the wave-function can written as
\begin{equation}
    \psi_{q_x}(x) = e^{iq_x x} u_q(x)  =  e^{iq_xx} \sum_n u_{G_n}(q_x+G_n) e^{iG_nx} =\sum_n u_{G_n}(q_x+G_n) e^{i(q_x + G_n) x} 
    \tag{S2}\label{S2}
\end{equation}
 and it follows that $ \psi_{q_x}(na) = e^{iq_x na} u_k(na)  = e^{iq_x na} u_q(0) =  e^{iq_x na} \psi_{q}(0) $. Also 
  $\psi_{q_x}(0) = \sum_n u_{G_n}(q_x+G_n)  $.
 Schr\"odinger's equation is then
\begin{equation}
\sum_n \left[ 
\frac{\hbar^2(q_x+ G_{n_x})^2 }{2m} - E_{q_x}
\right]  u_{G_n}(q_x +G_{n_x}) e^{i(q+ G_{n_x}) x} 
=
- V \psi_{q_x}(0)  \sum_n e^{i(q_x+G_n) x} 
\tag{S3}\label{S3}
\end{equation}
using  $ \sum_n \delta(x-a n) =  \sum_n e^{iG_nx}$ on the right hand side. The amplitude $u_{G_n}$ is extracted 
using $\frac{1}{L} \int^{L}_{0} e^{i(G^{\prime}-G)}dx =  \delta_{G^{\prime}G}$. Obtained is 
\begin{equation}
     \left[ \frac{\hbar^2(q_x+G)^2}{2m}- E_{q_x} \right] u_{G} = V \sum_{n}u_{G_{n}+G}  = V \sum_{n}u_{G_{n}} 
     \tag{S4}\label{S4}
\end{equation}
that reduces to
\begin{equation}
     u_{G_n} = {\cal G}_n (E_{q_x} )V
     \sum_{n}u_{G_{n}} 
     \tag{S5}\label{S5}
\end{equation}
where 
\begin{equation}
   {\cal G}_n (E_{q_x} ) = \frac{1}{E_{q_x} - \frac{\hbar^2(q_x + G_n)^2 }{2m}}  
   \tag{S6}\label{S6}
\end{equation}
is the free electron Green's function for momentum $\hbar (k_x + G_n)$. Summing over $n$ then gives
 \begin{equation}
     \frac{1}{V} =\sum_n  {\cal G}_n (E_{q_x} ) 
     \tag{S7}\label{S7}
 \end{equation}
 and a key result reproduced as Eq.~[1] in the main text. The equal weight of the ${\cal G}_n$ reflects the $\delta$-function nature of the periodic potential. The sum on $n$ is performed using $ \cot x = \sum_n \frac{1}{n\pi +x}$ to give the familiar result,
\begin{equation}
    \cos k_x a  =  \cos q_xa + \frac{\pi^2 V}{2E_0} \frac{\sin q_x a }{ q_xa}
    \tag{S8}\label{S8}
\end{equation}
and, for a given $q_x$, is solved for $k_x$. Since the potential is zero except where the delta-functions act, it can be insisted that the sought for energy $E_{q_x}(k_x) = \frac{{p_x}^2}{2m} = \frac{ \hbar^2}{2m}{ k_x}^2 $, where $p_x = \hbar k_x$, i.e., the $k_x$ in Eq.~\ref{S8} corresponds to the full momentum $p_x$ and not the crystal momentum $\hbar q_x$.  Here $E_0 = \frac{\pi^2 \hbar^2}{2ma^2}$  is the ground state energy of an isolated atom in one dimension. 

In the tight binding regime the bands develop from  ``atomic'' levels ($V \to \infty$) corresponding to 
\begin{equation}
    \sin q_x a = 0, \ \ \ \ \ \cos q_x a = (-1)^{n_x },\ \ \ \ q_x = \frac{\pi}{a} n_x; \ \ \ \ n_x=1,2, \ldots
    \tag{S9}\label{S9}
\end{equation}
Expanding about $q_xa = n_x\pi$, following some algebra, it is found
\begin{equation}
     E_{q_x}  = E_{n_x} - (-1)^n E_n  \frac{4E_0} {\pi^2 V}\cos q_x a 
     \tag{S10}\label{S10}
\end{equation}
and, as described in the text, the bands are centred about the atomic levels $E_{n_x} = {n_x}^2 E_0$. These results are illustrated in Fig.~[1a] of the main text.

In two dimensions and a total potential $V(x,y,z) = V(x)+ V(y)$, with the same definition of $V(x)$, the wave function $\Psi_{\vec k} (x,y,z) = \psi_{k_x}(x) \psi_{k_y}(y) $  is separable. There are two separation constants $E_{k_x}$ and   $E_{k_y}  $ and the full 
  Schr\"odinger equation 
\begin{equation}
    \left[
 - \frac{\hbar^2}{2m} \nabla^2 + V(x,y,z) 
 \right]
\Psi_{\vec q}  (x,y,z)  = E_{\vec q } \Psi_{\vec q}  (x,y,z) 
\tag{S11}\label{S11}
\end{equation}
reduces to Eq.~\ref{S1} for $\psi_{q_x}(x)$ with equivalent equations for $ \psi_{q_y}(y)$. The constants of separation add to give the energy
\begin{equation}
    E_{\vec q }  =  E_{q_x}  +  E_{q_y} 
    \tag{S12}\label{S12}
\end{equation}
the simple sum of the energies for the two degrees of freedom. 

Since the model is separable, the tight binding regime in two dimensions corresponds to ``atomic'' levels $E_0  ({s_x}^2+ {s_y}^2)  $ characterised by 
 quantum numbers $s_x$ and $s_y$. Adding the one dimensional energies:
\begin{equation}
    E_{q_x q_y }  = E_0  ({s_x}^2+ {s_y}^2)    +  \frac{4E_0} {\pi^2 V}
 [ (-1)^{n_x} E_{s_x} \cos q_x a + y(-1)^{n_y} E_{s_y} \cos q_y a ]
 \tag{S13}\label{S13}
\end{equation}
The resulting two dimensional band structure is illustrated in Fig.~[1b].
The similar three dimensional band structure develops from the ``atomic'' levels $E_0  ({s_x}^2+ {s_y}^2+{s_z}^2)  $ as discussed in the text. 

{\it The  magnetic Kronig-Penny model
}
Since the thin film is magnetic, added to the Hamiltonian is the Stoner exchange interaction $- J_{0}\vec{S}\cdot \vec{\sigma} $ where $\vec S$ is the magnitude of the order parameter, i.e., the effective spin that is determined self-consistently. This might be imagined to be a simple version of the exchange correlation energy in the DFT. The density matrix corresponds to a magnetisation $\vec M$ pointing in the direction of $\vec S$. Assuming $\vec M = M\hat{\bf z}$, the basic Eq.~\ref{S7} is unchanged except the diagonal Green's function 
\begin{equation}
    {\cal G}_n^{-1}  (E_{q_x} ) = [(E_{q_x}-\frac{\hbar^2(q_x+G_n)^2}{2m})I+ J_{0}S \sigma_{z}].
    \tag{S14}\label{S14}
\end{equation}
is now a two-by-two matrix. The Mathematica solutions of this are described in the text in the connection with Fig.~[4a] . The solutions are shifted up or down in energy by $J_0 S$ as the band corresponds to minority/majority electrons.

{\it The  periodic Rashba model.
}
This illustrates the elementary diagonalisation of the spin problem by a rotation matrix. 
 In two dimensions, the spin-orbit interaction $ \frac{\alpha_{R}}{\hbar}(\sigma_{x}p_{y}-\sigma_{y}p_{x})$ is added to the Hamiltonian.  Unfortunately, because of the Pauli matrices $\sigma_{x}$ and $\sigma_{y}$ in the Rashba term, this is no longer exactly separable in two dimensions. In a single dimension the chains are  taken to be the $x$-direction implying a spin-orbit coupling $\frac{\alpha_{R}}{\hbar}\sigma_{y}p_{x}$.  The problem still reduces to Eq.~\ref{S7} but where now the free electron propagator:
\begin{equation}
    {\cal G}_{n_x}^{-1}  (E_{q_x} ) = [(E_{q_x}-\frac{\hbar^2(q_x+G_{n_x})^2}{2m})I+\alpha_{R}(q_x+G_n)\sigma_{y}],
    \tag{S15}\label{S15}
\end{equation}
again a two-by-two matrix but now with off-diagonal matrix elements.
This can be made spin diagonal by a rotation of $\frac{\pi}{2}$ about the $x$-axis.  The rotation matrix $R_{x}(\frac{\pi}{2})$ is such that $\sigma_y = (\frac{\pi}{2})\sigma_{z}R_{x}(\frac{\pi}{2})$. It follows $[(E_{q_x}-\frac{\hbar^2(q_x+G_{n_x})^2}{2m})I+ \alpha_{R}(q_x+G)R_{x}^{\dag}(\frac{\pi}{2})\sigma_{z}R_{x}(\frac{\pi}{2})] =R_{x}^{\dag}(\frac{\pi}{2}) [(E_{q_x}-\frac{\hbar^2(q_x+G_{n_x})^2}{2m})I+\alpha_{R}(q_x+G_{n_x})\sigma_{z}]R_{x}(\frac{\pi}{2})$. Since 
$ R_{x}^{}(\frac{\pi}{2})$ is independent of $n_x$, these rotation matrices can be factored out of the sum on $n_x$, i.e., 
\begin{equation}
    \frac{1}{V} =R_{x}^{}(\frac{\pi}{2}) \left( \sum_{n_x}  {\cal G}_{n_x} (E_{q_x} ) \right) R_{x}^{\dag}(\frac{\pi}{2}) \to  \frac{1}{V} =\sum_{n_x}   {\cal G}_{n_x} (E_{q_x} )
    \tag{S16}\label{S16}
\end{equation}
multiplying on the left by $R_{x}^{\dag}(\frac{\pi}{2}) $ and the right by $R_{x}^{}(\frac{\pi}{2}) $. 
(The same diagonal result is obtained trivially be using a representation of the Pauli matrices in which $\sigma_y$ is directly diagonal.) The spin dependent momentum shift $\alpha_{R}(k+G_{n_x} )\sigma_{z}$ is absorbed into the kinetic energy, i.e., 
 $ [(E_{q_x}-\frac{\hbar^2(q_x+G_{n_x} )^2}{2m})I+ \alpha_{R}(q_x+G_{n_x})\sigma_{z}] =
 [(E_{q_x}-\frac{\hbar^2(q_x+G_{n_x} )^2}{2m})I\pm \alpha_{R}(q_x+G_{n_x} )] = [(E_{q_x}- \frac{\hbar^2(q_x+G_{n_x} \mp\Delta k)^2}{2m} - E_R]$ upon completing the square. The band structure is shifted to the left or right by $\Delta k = \frac{m}{\hbar^2}{\alpha_R}$ and down in energy by
\begin{equation}
    E_R = \frac{\hbar^2{\Delta k}^2}{2m}  = \frac{1}{2}m \frac{{\alpha_R}^2} {\hbar^2}
    \tag{S17}\label{S17}
\end{equation}
This is the same  characteristic energy identified for the free electron case\cite{Barnes2014}, i.e., when $V =0$. However the identification of an effective $``E_R"$ for the different bands is changed as discussed in the text in the context of Fig.~[4b]. The determination of $\alpha_R$ from the experimental $``E_R"$ cannot be justified.

{\it The  two dimensional periodic magnetic Rashba model.}  Now in two dimensions, both the spin-orbit interaction $ \frac{\alpha_{R}}{\hbar}(\sigma_{x}p_{y}-\sigma_{y}p_{x})$ and exchange term $- J_{0}\vec{S}\cdot \vec{\sigma} $ are added to the Hamiltonian.   Without the periodic potential, this is the model considered by Barnes et al\cite{Barnes2014}. With the periodic potential the Hamiltonian is therefore
\begin{equation}
    \hat{H} = \frac{p^2}{2 m} - J_{0}\vec{S}\cdot \vec{\sigma} + \frac{\alpha_{R}}{\hbar}(\sigma_{x}p_{y}-\sigma_{y}p_{x}) + V [\sum_{{G_{n}}_x } e^{-i {G_{n}}_x x}
 +\sum_{{G_{n}}_y } e^{-i {G_{n}}_y y}.
 ]
 \tag{S18}\label{S18}
\end{equation}
The model still has both free electron and tight binding limits. Again, because of the Pauli matrices $\sigma_{x}$ and $\sigma_{y}$ in the Rashba term, this is no longer {\it exactly\/} separable in two dimensions. How this can be approximately separated in two or higher dimensions will be taken up again below.

{\it The  one dimensional periodic magnetic Rashba model - in plane
} 
In a single dimension,  first imagine, the equivalent to the two dimensional ``in-plane'' case, see Fig.~[4c]. The chain is taken to be the $x$-direction implying a spin-orbit coupling $\frac{\alpha_{R}}{\hbar}\sigma_{y}p_{x}$. In-plane here means that $\vec S = S\hat {\bf y}$, i.e., the exchange $\vec B_E$ and Rashba field $\vec B_R$ are parallel and both involve only $\sigma_{y}$. Again Eq.~\ref{S7} applies but now with the matrix
\begin{equation}
    {\cal G}_n^{-1}  (E_{q_x} ) = [(E_{q_x}-\frac{\hbar^2(q_x+G_{n_x})^2}{2m})I+\left(J_{0}S +\alpha_{R}(q_x+G_{n_x})\right)\sigma_{y}].
    \tag{S19}\label{S19}
\end{equation}
 As with the non-magnetic case, this can be made spin diagonal by a rotation of $\frac{\pi}{2}$ about the $x$-axis.
 Now in Eq.~~\ref{S7} the diagonal Green's function is 
\begin{equation}
    {\cal G}_n^{-1}  (E_{q_x} ) = [(E_{q_x}-\frac{\hbar^2(q_x+G_{n_x})^2}{2m})I+\left(J_{0}S +\alpha_{R}(q_x+G_{n_x})\right)\sigma_{z}].
    \tag{S20}\label{S20}
\end{equation}
Comparing Eq.~\ref{S15} with Eq.~\ref{S20}, little changes as compared to the non-magnetic Rashba problem except the  minority/majority shifts band shifts of $\pm J_{0}S$ in addition to those of the simpler problem. Again the solutions exhibited in the text are 
 obtained with Mathematica.
 
 It is a key observation that the shift in energy due to the Rashba effect is $E_R$ for all bands and is second order in the electric field $E$. This is perhaps surprising since the Rashba field itself $B_R = \alpha_{R}(q_x+G_{n_x})$ is linear in $E$. The momentum shifts are indeed linear, however the process of completing the square reduces the energy shift to second order.

{\it The  one dimensional periodic magnetic Rashba model - out of plane
} 
Turning to the more interesting one dimensional version of an ``out-of-plane'' magnetisation, i.e., when the average order parameter  $\vec M = M\hat {\bf z}$ is parallel to the direction of the electric  field $\vec E = E\hat {\bf z}$. The combination $-J_0S \sigma_z + \frac{\alpha_{R}}{\hbar}\sigma_{y}p_{x}$ is diagonalised by a rotation $R_{x}(\theta_G)$ such that $R_{x}(\theta_G){J}S\sigma_{z}R^{\dag}_{x}(\theta_{G}) = J_{0}S\sigma_{z} +\alpha_{R}(q_x +G_{n_x})\sigma_{y}$ where there is a new effective exchange
\begin{equation}
    SJ = \sqrt{J_{0}^{2}S^2 + \alpha_{R}^{2}(k+G)^2} \approx J_0 S + \frac{\alpha_{R}^{2}(q_x+G_{n_x})^2}{2J_0S}
    \tag{S21}\label{S21}
\end{equation}
 and $ \theta_{G_{n_x} } = \arctan(\frac{\alpha(q_x +G_{n_x})}{J_{0}S})$. The Rashba field $B_R = \alpha_{R}(q_x+G_{n_x})$ is unchanged in magnitude, but is now  perpendicular to the exchange field $B_E =  J_0 S$, and  inevitably, via vector addition, leads to a second order, i.e., $E^2$, correction assuming $B_E$  is the larger field. As explained in the previous section, the in-plane magnetic anisotropy energy is also second order in $E$. This  is the reason for the counter intuitive result that a Rashba field perpendicular the magnetisation can lead to a greater magnetic anisotropy energy that does the in-plane field of equal magnitude. Schr\"odinger's equation reduces to 
\begin{equation} 
 \frac{a}{V} =\sum_{n_x}R_{x}^{\dag}(\theta_{G_{n_x}})
           [ (\frac{1}{2} (1+\sigma_z) {\cal G}^{+}_{G_{n_x}}+\frac{1}{2} (1-\sigma_z){\cal G}^{-}_{G_{n_x}})  R_{x}(\theta_{G_{n_x}})
           \tag{S22}\label{S22}
\end{equation}
 where now the free electron Green's functions $ {\cal G}^{\pm}_{n_x} = [E_{q_x} -\frac{\hbar^2(q_x+G_{n_x})^2}{2m}\pm {JS}]^{-1}$ includes the exchange splitting with the new $JS$ that includes the second order Rashba correction $ \frac{\alpha_{R}^{2}(q_x+G_{n_x})^2}{2J_0S}$. While the Green's function has been reduced to diagonal form, this has been at the cost of introducing the non-trivial rotation matrices $R_{x}(\theta_{G_n})$ that can no longer be factored out of the sum and eliminated.  
 
 The appearance of $\sigma_z$ in combination with the rotations $ R_{x}(\theta_{G_{n_x}})$ implies a non-trivial Berry phase. The quantity $ \sigma_z \equiv e^{i \frac{\pi}{2}(1- \sigma_z)} = e^{i \frac{\pi}{2} } e^{ - i \frac{\pi}{\hbar}  s_z} $, and apart from the phase factor $ e^{i \frac{\pi}{2} } $, corresponds to a rotation by $\pi $ about the $\hat{\bf z}$-axis. It follows:
 \begin{equation}
R_{x}^{\dag}   \frac{1}{2} (1\pm\sigma_z)  R_{x} =   \frac{1}{2}  (1  \pm R_{x}^{\dag}  e^{i \frac{\pi}{2} }  e^{i \pi s_z/\hbar} R_{x}) %=   \frac{1}{\sqrt{2}}  e^{\pm i\frac{\pi}{4}\sigma_z} 
=   \frac{1}{2}  (1  \pm   e^{i \frac{\pi}{2} }  e^{-i  \frac{\pi}{\hbar}  (\cos \theta_{G_{n_x}} s_z + \sin \theta_{G_{n_x}} s_y)}).
      \tag{S23}\label{S23}
\end{equation}
As with the Stern-Gerlach\cite{SternGerlach} experiment, an electron propagates along two similar paths that interfere with each other. The electrically controlled spin Berry phase corresponds to a ``magnetic flux" enclosed by these paths and for small fields is second order in the electric field $E$. However, the  off-diagonal part $- \frac{\pi}{\hbar}\sin \theta_{G_{n_x}} s_y$ implies linear in $E$ spin-mixing that leads directly to electrically controlled band-structure gaps.
 These predicted effects provide a novel experimental possibility  in which spin-Berry phase effects can be explored by electrical gating of thin film magnets.

In a concise form, the equivalent of Schr\"odinger's equation is therefore
\begin{equation}
     \frac{1}{V} =
             \frac{1}{2} \sum_\pm \sum_{n_x}
            %e^{\pm i\frac{\pi}{4}(\cos \theta_{G_{n_x}} \sigma_z + \sin \theta_{G_{n_x}} \sigma_y)}  
            (1  \pm   e^{ -i \frac{\pi}{2} (\cos \theta_{G_{n_x}} +\sigma_z)\sigma_z - \sin \theta_{G_{n_x}} \sigma_y) } )
            {\cal G}^{\pm}_{G_{n_x}}
             \tag{S24}\label{S24}
\end{equation}
as reproduced as Eq.~[2] in the text. 
In the phase factor $
 e^{- i  \frac{\pi}{\hbar}  (\cos \theta_{G_{n_x}} s_z + \sin \theta_{G_{n_x}} s_y)}$,
the quantity $\sigma^\prime_z = (\cos \theta_{G_{n_x}} \sigma_z + \sin \theta_{G_{n_x}} \sigma_y)$ is such that $(\sigma^\prime_z)^2 = 1$ and it follows
\begin{equation}
  e^{i\frac{\pi}{2} } e^{-i\frac{\pi}{2} \sigma^\prime_z}  =  \sigma^\prime_z = \cos \theta_{G_{n_x}} \sigma_z + \sin \theta_{G_{n_x}} \sigma_y
 \tag{S25}\label{S25}
\end{equation}
and hence, following  some algebra, Eq.~\ref{S24} reduces to 
\begin{equation}
    \frac{1}{V} = {\cal S} +  {\cal A}  \sigma_{z}  + {\cal B}  \sigma_{y}
    \tag{S26}\label{S26}
\end{equation}
 where ${\cal S} \equiv  \frac{1}{2}\sum_{n_x}({\cal G}^{+}_{G_{n_x}}+{\cal G}^{-}_{G_{n_x}})$,  ${\cal A} \equiv  \frac{1}{2}\sum_{n_x}({\cal G}^{+}_{G_{n_x}}-{\cal G}^{-}_{G_{n_x}})\cos{\theta_{G_{n_x}}}$ and  ${\cal B} \equiv  \frac{1}{2}\sum_{n_x}({\cal G}^{+}_{G_{n_x}}-{\cal G}^{-}_{G_{n_x}})\sin{\theta_{G_{n_x}}}$. 
 Following yet another spin rotation by the angle $\Theta= \arctan{\frac{\cal B}{\cal A}}$ the majority/minority  spin-split bands are given by the solutions of  $ \frac{1}{V} - {\cal S} +  \sqrt{{\cal A}^{2}+{\cal B}^{2}} \sigma_z =0 $ as $\sigma_z=\pm1$. Finally, performing a square to eliminate $ \sigma_z $, results in 
\begin{equation}
%\label{111X}
        \left(\frac{1}{V}-{\cal S}\right)^2 
        - \left({\cal A}^2
        +{\cal B}^2\right)=0
        \tag{S27}
        \label{S27}
    \end{equation}
an implicit equation for $E_{q_x}$, including spin mixing effects, that is fully equivalent to the original Schr\"odinger equations for both spin directions. Both of the majority and minority bands are obtained by Mathematica solutions of this single equation.

{\it Spin Berry physics.
} 
From Ref.~[11], with the present definitions of the rotation matrices, the spin Berry connection is
\begin{equation}
    \vec A_s = -\frac{\hbar}{2e}(\sin \theta_{G_{n_x}} \vec \nabla \phi \sigma_y + \vec \nabla \theta \sigma_x)
    \tag{S28}\label{S28}
\end{equation}
The somewhat generalised spin Berry phase $\phi_s$ is then defined by
\begin{equation}
    e^{i\phi_s} = :e^{i\frac{e}{\hbar}  \int_c  \vec A_s \cdot d\vec r }: %= e^{\pm i\frac{\pi}{4}(\cos \theta_{G_{n_x}} \sigma_z + \sin \theta_{G_{n_x}} \sigma_y)} 
    \tag{S29}\label{S29}
\end{equation}
where the $:\ldots :$ implies spatial ordering. The spatial ordering is important since the $\sigma_x$ and $\sigma_y$ that occur in the definition of $\vec A_s$  do not commute. In general $\phi_s$ is an operator written as a linear combination of Pauli matrices. There is a magnetic monopole at the centre of the spin sphere, compensated by a Dirac string. The contour over which the integration in Eq.~\ref{S29} is preformed is dictated by the 
 rotations 
\begin{equation}
    R_{x}^{\dag}(\theta_{G_{n_x}}) 
     %e^{i \frac{\pi}{2} } 
     e^{  i \frac{\pi}{\hbar}(1-s_z)} 
     %e^{\pm i\frac{\pi}{2} \sigma_z} 
     R_{x}(\theta_{G_{n_x}}).
    \tag{S30}\label{S30}
\end{equation}
For an initial up spin, this path is  illustrated  in the text Fig.~[1g]. A  beginning up spin is first rotated down by the angle $\theta_{G_{n_x}}$ then about the $\hat {\bf z}$-axis by $\pi$ and finally back to its original alignment along the $\hat {\bf z}$-direction. Because of the factor of $e^{i \frac{\pi}{2} }$, the solid angle $\Omega$ is defined relative the the positive $\hat {\bf z}$-axis for up-spin and the negative axis for down-spin. 
That part of $:e^{i\frac{e}{\hbar}  \int  \vec A \cdot d\vec r }:$ that arises from the commutators of $\sigma_x$ and $\sigma_y$ is given by a generalised Stoke's theorem. The proof is straightforward[11] and will not be given here since it leads to the well known result that the diagonal Berry phase is $\pm \frac{1}{2} \Omega$ where $\Omega$ is the solid angle subtended by the path on the Bloch sphere relative to the positive or negative $\hat {\bf z}$-axis as the spin is up or down. A familiar result\cite{Shankar1994}  is the the solid angle for a full circle about the positive $\hat {\bf z}$-direction is $\Omega = 2\pi (1-\cos \theta)$. For half a circle the (diagnal) Berry phase is therefore
\begin{equation}
    \frac{\pi}{2}(1- \cos \theta_{G_{n_x}})\sigma_z.
    \tag{S31}\label{S31}
\end{equation}
 If the commutators are ignored, the spatial ordering is not important and the remaining part of the  phase is given directly by the integral 
\begin{equation}
    \frac{1}{2} \int_c (\sin \theta_{G_{n_x}} d \phi  \sigma_y + d \theta  \sigma_x) =  \frac{\pi}{2}\sin \theta_{G_{n_x}}\sigma_y.
    \tag{S32}\label{S32}
\end{equation}
since the $\int_cd \theta = 0$ by symmetry. This is intended as merely a sketch of how the general definition[11], Eq.~\ref{S28}, that defines $\vec A_s$, leads to the same result as the algebraic manipulations that lead to Eq.~\ref{S24}, i.e., the Berry phase factor 
\begin{equation}
    e^{i\phi_s} = :e^{i\frac{e}{\hbar}  \int  \vec A \cdot d\vec r }: %= e^{ i\frac{\pi}{4}(\cos \theta_{G_{n_x}} \sigma_z + \sin \theta_{G_{n_x}} \sigma_y)} 
    =e^{ -i \frac{\pi}{2} (\cos \theta_{G_{n_x}} +\sigma_z)\sigma_z + \sin \theta_{G_{n_x}} \sigma_y) } 
    \tag{S33}\label{S33}
\end{equation}
for the appropriate spin up/down contour. When compared with the standard approach to the Berry phase, this algebraic definition[11] of the real space connection $\vec A_s(\vec r)$, leads in a natural manner, to the off-diagonal part of the Berry phase factor $e^{i\phi_s}$ with the on-diagonal corresponding to a curvature defined by a Stoke's theorem. It is however the off-diagonal part that leads to the electrical control of spin gaps as described in the text. This prediction is clearly accessible to experimentation.

{\it Out of plane - tight binding - analytic results.
} 
Analytic results for the tight binding regime are obtained by using the approximation $SJ \approx J_0S
+  \alpha_{R}^{2}(q_x+G)^2/J_0S$ valid given  the Rashba field $ \alpha_{R}\hbar (q_x+G_{n_x} )$ is small when compared with the exchange field $J_0S$. It follows $ E^\pm_n(k) \approx \frac{\hbar^2(q_x+G_{n_x} )^2 }{2m} \mp (J_0S  + {\alpha_{R}}^{2} \frac{\hbar^2 (q_x+G_{n_x} )^2}{2 J_0S}  )$. 
The Rashba correction $ \mp  {\alpha_{R}}^{2} \frac{\hbar^2(q_x+G_{n_x} )^2}{2J_0 S} $ is then absorbed into a spin dependent  effective mass $\mu$ defined by $\frac{1}{\mu^\pm} = \frac{1}{m} \mp   \frac{{\alpha_{R}}^{2} }{J_0 S}  $. The centre of tight binding band with index $n_x$ is then $E^\pm_n = n^2 \frac{ \pi^2 \hbar^2}{ 2\mu a^2 } = E_n \mp  \frac{ 1}{  4 } {G_{n_x}}^2 \frac{\hbar^2{\alpha_{R}}^{2} }{2J_0 S}$ using $\frac{1}{2} G_n = n \frac{\pi}{a}$. The resulting  energies  are
\begin{equation}
E^\pm_{q_x} = E_{n_x}-  \frac{4E_0E_n } {V}\cos q_xa  \mp \left(  J_0 S +  {\alpha_{R}}^{2} \frac{\hbar^4 }{8J_0 S}   {G_{n_x}}^2 \right)
     \tag{S34}\label{S34}
\end{equation}
for the majority/minority bands. In the text, this result is discussed in the  context of  the  Fig.~[4d]. The role of the reciprocal lattice vectors in determining the Rashba field $B_R$ is put in evidence. 

{\it Electrically controlled spin gaps - spin mixing effects %- the spin Berry phase
} 
If neither $\alpha_R$ or  $V$ have extreme values, then it is possible to develop an approximate theory for the interaction between bands by reducing the sum over Green's functions in Eq.~\ref{S7} to only two terms. The secular equation reduces to 
\begin{equation}
\frac{1}{V} = \sum_{n_x}  G_{n_x} (E_{k_x} ) = G_a (E_{q_x} ) + G_b (E_{q_x} ).
     \tag{S35}\label{S35}
\end{equation}
  To be specific consider the first such gap near $k_x = \frac{\pi}{a}$.  The Green's function for the lowest band $G_0 = [E_{k} -\epsilon_c + \frac{\hbar^2 }{2m}\frac{2\pi}{a} (q-k_c)]^{-1}$ where $\epsilon_c \approx  \frac{\hbar^2 }{2m}(\frac{\pi}{a})^2$ is the energy where the two free electron bands cross. The other function $G_{2\pi/a}  = [E_{k} -\epsilon_c - \frac{\hbar^2 }{2m}\frac{2\pi}{a} (q-k_c)]^{-1}$ corresponds to the free electron band translated from beyond $-\frac{\pi}{a} $ to near $k_x = \frac{\pi}{a}$. Since the free electron bands cross at $\epsilon_c$ the Green's functions differ only in the direction of dispersion, i.e., in the sign of the term $\Delta = \frac{\hbar^2 }{2m}\frac{2\pi}{a} (q-k_c)$. The Rashba and exchange fields are contained in the definition of $  k_c = \frac{2m}{\hbar^2 }\frac{a}{2\pi} J S \approx  \frac{2m}{\hbar^2 }\frac{a}{2\pi} (J_0S  + {\alpha_{R}}^{2} \frac{\hbar^2 {{k_F}^2}}{2 J_0S}  )$. The Rashba effect is clearly $O( {\alpha_{R}}^{2} )$ in the absence of 
the spin mixing rotations $\theta_0\approx - \theta_{2\pi/a}   \approx  \frac{\alpha_R k_c}{J_0S}\equiv \theta $. Substituting the definitions of $\cal S$, $\cal  A$ and $\cal  B$ into $\frac{1}{V} = {\cal  S} +  {\cal  A}  \sigma_{z}  + {\cal B}  \sigma_{y}$, following some involved algebra, obtained for the energies is
\begin{equation}
E_{k} =\epsilon_c+ V+\sigma_z\frac{V}{2}\sqrt{\left( \frac{\alpha_R\pi}{J_0S a}\right)^2+\left(\frac{\hbar^2 }{m}\frac{\pi}{V} \right)^2(q-k_c)^2}
    \tag{S36}\label{S36}
\end{equation}
The two branches near $ \epsilon_c +  V $ disperse up or down as $ \sigma_z  = \pm 1$. The key result is that the gap evident in Fig.~[4e] is approximately 
\begin{equation}
\left|V\frac{ 1} {J_0S }(\alpha_Rk_c)\right| 
\tag{S37}\label{S37}
\end{equation}
and is linear in the  Rashba parameter $\alpha_R$ and hence the electric field $E$.

{\it Electrically controlled spin gaps - measuring the spin Berry phase.
} 
The more general result for the gap is 
\begin{equation}
\left|V\theta_{G_{n_x}}\right|\tag{S38}\label{S38}
\end{equation}
where the angle $\theta_{G_{n_x}}$ reflects a unitary rotation matrix $R_{x}(\theta_{G_{n_x}})=e^{-i\theta_{G_{n_x}}s_x/\hbar}$ and where $n_x$ is the band index of the lower of the two bands. This rotation can be traced back to the spin dependent geometrical phase $e^{-i\theta_{G_{n_x}}s_x/\hbar}$ that appears in the wave-function
\begin{equation}
\psi_{q_x}(x) = \sum_n e^{-i\theta_{G_{n_x}}s_x/\hbar}|u(q_x+G_{n_x})| e^{i(q_x+G_{n_x} ) x} 
    \tag{S39}\label{S39}
\end{equation}
This causes a spin dependent Aharonov–Bohm (or Stern-Gerlach) type interference between the different amplitudes $|u(q_x+G_{n_x} )| $ as the electron is reflected by the periodic potential. Because the momentum $\hbar(q_x+G_{n_x} )$ changes sign, so 
does  $\theta_{G_{n_x}}$ while as usual for a hard potential there is a $\pi$ change in the phase $\phi$. 
The angles $\theta_{G_{n_x}}$ can be controlled by the electric field $E$ through $\alpha_R$ and hence dictated  by experiment. Such measurements permit the determination of the Berry phase factor $e^{-i\theta_{G_{n_x}}s_x/\hbar}$ and again
the existence of these spin-gaps provides a novel test of spin  Berry phase physics.

{\it Approximate separation of variables in two plus dimensions.
} 
At first sight, the two dimensional Rashba term  $\frac{\alpha^y_{R}}{\hbar} \sigma_{x}p_{y} - \frac{\alpha^x_{R}}{\hbar}\sigma_{y}p_{x} $ is separable since, e.g., $ \frac{\alpha^y_{R}}{\hbar} \sigma_{x}p_{y} $ only operates on functions of $y$. This is not the case since the spin wave function, upon which $\sigma_x$ and $\sigma_{y}$ act, cannot be factored. This not withstanding, the problem is indeed separable if the rotation angles  $\theta_{G_{xn}}(q_x) $ defined earlier by 
$R_{x}(\theta_{G_{xn}}(q_x) ){J}S\sigma_{z}R^{\dag}_{x}(\theta_{_{G_{n_x}}}(q_x) ) = J_{0}S\sigma_{z} +\alpha^x_{R}(q_x+G_{n_x} )\sigma_{y}$
are small. In this case it is  possible to write 
\begin{align*}
&R_{x}(\theta_{G_{n_x}}(k_x) ){J}S\sigma_{z}R^{\dag}_{x}(\theta_{_{G_{n_x}}}(k_x) )+
R_{y}(\theta_{G_{n_y}}(k_y) ) {J}S\sigma_{z}R^{\dag}_{y}(\theta_{G_{n_y}}(k_y) )   - {J_0}S\sigma_{z}\\
 &= J_{0}S\sigma_{z} +\alpha^x_{R}(q_x+G_{n_x})\sigma_{y}-\alpha^y_{R}(k_y+G_{n_y} )\sigma_{x}.
 \tag{S40}\label{S40}
\end{align*} 
A separable solution is of the form: $E_{q_x, q_y} \psi_x (x)  \psi_y(y ) |\rangle = H  \psi_x (x)  \psi_y(y ) |\rangle $, where $ |\rangle $ is the spin ket. This reduces to 
\begin{equation}
\Big[ \frac{1}{\psi_x (x) }\hat H_x \psi_x (x)+\frac{1}{\psi_y (y) } \hat H_y\psi_y (y) 
\Big] |\rangle=(E_{q_x, q_y} +  SJ_{0}\sigma_z )  |\rangle
\tag{S41}\label{S41}
\end{equation}
where, $ \hat{H}_x$ was given earlier and $ \hat{H}_y = (p^2/2 m) -R_{y}(\theta_{G_{n_y}}(q_y) ){J}S\sigma_{z}R^{\dag}_{y}(\theta_{_{G_{n_y}}}(q_y) ) + V(y) $. The right hand side of this equation is an, $x$ and $y$ independent, two-by-two matrix acting on the Pauli spinor $ |\rangle$ and the same is true for both terms on the left hand side. It must be that, e.g., 
\begin{equation}
\frac{1}{\psi_x (x) } [ \frac{1}{2m}  {p_x}^2 -R_{x}(\theta_{G_{xn}}(q_x)){J}S\sigma_{z}R^{\dag}_{x}(\theta_{_{G_{xn}}}(q_x) )+V(x)] \psi_x (x)= M_x(q_x)
\tag{S42}\label{S42}
\end{equation}
where the matrix 
$ M_x(q_x) = E^x(q_x) {\bf I} - J^x (q_x) S \sigma_z - \alpha^x_R e^x(q_x) \sigma_y $
and where $E^x(q_x) $, $J^x (q_x) $ and $e(q_x)$ are three separation constants. That $ M_x(q_x) $ involves only $\sigma_y$ and not $\sigma_x$ reflects the Rashba term $ \frac{\alpha^x_{R}}{\hbar}\sigma_{y}p_{x}$. Given the  $M_x(q_x) $ and $ M_y(q_y) $ are determined, there then remains the spin Schr\"odinger's equation:
\begin{equation}
[(M_x(q_x)+ M_y(q_y))-(E_{q_x, q_y} +  SJ_{0}\sigma_z)]|\rangle = 0
\tag{S43}\label{S43}
\end{equation}
that is trivially solved by a rotation in spin space. The resulting spin ket determines $q_x$, $q_y$ dependent density matrix but leaves the Berry phase angles, e.g., $\theta_{G_{n_x}}$  hidden within  $M_x(q_x)$. Since there are many solutions for a given crystal momentum $\vec q$ there are equally many different separation matrices $M_x(q_x)$.

The above for  $\psi_x (x) $ can be re-arranged to read
\begin{equation}
\frac{1}{\psi_x (x)}[\frac{1}{2m}{p_x}^2-E^x(q_x)+
-S(J_{0}-\Delta E)\sigma_z+\alpha^x_{R} \sigma_{y}(\frac{p_{x}}{\hbar}-e)]\psi_x(x)=V(x)
\tag{S44}\label{S44}
\end{equation}
where $E^x(q_x) = \frac{1}{2} (E_\uparrow^x(q_x) + E_\downarrow^x(q_x) )$ and $\Delta E = \frac{1}{2} [(E_\uparrow^x(q_x) - E_\downarrow^x(q_x) ]$. This is of the original, one dimensional, form but with the extra parameters $E^x(q_x) $ and $\Delta E$ that modify the exchange and Rashba terms. Following the same steps as before,  the problem for $ M_x(q_x) $ reduces to  
\begin{equation}
( \frac{1}{V_x} + {\cal S}_x + {\cal A}_x \sigma_z + {\cal B}_x \sigma_y) = 0
\tag{S45}\label{S45}
\end{equation}
but with the same  definitions of ${\cal S}_x$, ${\cal A}_x$ and  ${\cal B}_x$ in terms of
\begin{equation}
\theta_{G_{n_x}} = \arctan(\frac{\alpha^x(q_x+G_{n_x} )-e(q_x)}{J_{0}S}).
\tag{S46}\label{S46}
\end{equation}
The Green's functions are again
\begin{equation}
{ \cal G}^{\pm}_{G_{n_x}} = \frac{1}{E^x_{q_x} -\frac{\hbar^2(q_x+G_{xn})^2}{2m}\pm {J}S}.
\tag{S47}\label{S47}
\end{equation}
but with the effective exchange including Rashba corrections and the parameter $ \Delta E$:
\begin{equation}
SJ =  \sqrt{J_{0}^{2}S^2 + {\alpha^x_{R}}^{2}\hbar^2 (q_x+G_{xn})^2 }- \Delta E
\tag{S48}\label{S48}
\end{equation}
Different at this point is that  the coefficients of $\sigma_z$, $\sigma_y$ are individually required to be zero, i.e., it is insisted ${\cal A}_x =0$ and  ${\cal B}_x =0$,  leaving $ \frac{a}{V_x} + {\cal S}_x  = 0$. These three equations yield multiple solutions for the three separation constants  $E^y(q_y) $, $J^y (q_y) $ and $e(q_z)$ that determine $M_y(q_y)$.  With similar equations for $\psi_y (y)$, the problem reduces to the spin Schr\"odinger's equation given above. In principle, these sets of equations reduce the problem to quadrature but in practice, given the many solutions for $E^y(q_y) $ they overwhelm Mathematica. They can be solved in simple limits as described in the text. Extension of these results by limiting the number of terms in the sums on $n_x$ and $n_y$ is possible but beyond the work presented here. The method can evidently extended to higher dimensions

%Iwasaki, Y., Hashimoto, Y., Nakamura, T. et al. Conductance fluctuations in InAs quantum wells possibly driven by Zitterbewegung. Sci Rep 7, 7909 (2017). https://doi.org/10.1038/s41598-017-06818-4

{\it Spin-orbit coupling
} 
The full spin-orbit coupling   can be written with a number of ways:
$$
V_{\rm so} 
= - \frac{e\hbar }{4m^2 c^2}\vec E \cdot  (\vec \sigma\times \vec p)
= -  \frac{ \lambda_c}{8\pi mc}  \vec E \cdot  (\vec \sigma\times \vec p)
= - \mu_B \frac{1}{2c^2}  \vec E \cdot  (\vec \sigma\times \vec v)
= \frac{\alpha_R}{\hbar}\vec  \sigma \cdot (\vec p \times \hat{\bf z})
$$
where the first version illustrates $V_{\rm so} $ arises from an expansion in the mass gap $\frac{1}{mc^2}$, the second relates this expansion to the Compton wave-length $\lambda_c =\alpha a_0\approx 2$pm, $\alpha \approx \frac{1}{137} $ the fine-structure constant and the expansion parameter of QED and where $a_0 \approx 0.5$\AA\ is the Bohr radius. The next version re-enforces the idea that the Rashba field $B_R =  \frac{v}{2c^2} E$ is just the magnetic field seen in the electron's frame due to an electric field in the laboratory frame and where the factor of $\frac{1}{2}$ accounts for the Thomas rotation. The last version puts in evidence that the $V_{\rm so} $ is of the form of a Rashba spin-orbit coupling. This is the entire 
 such coupling if $\vec E = E \hat{\bf z}$ is restricted to the applied electric field and/or the internal field that occurs in the absence of an inversion symmetry. The apparently subtle point being made here is that this last version involves the full momentum $\vec p = \hbar \vec k$ while the usual Rashba term as derived by Rashba and co-workers\cite{Rashba1984} or described in the exhaustive work found in Winkler's text\cite{Winkler2003}, is rather the crystal momentum $\hbar \vec q$ that occurs in the triple product $\vec  \sigma \cdot (\vec p \times \hat{\bf z})$, where the wave-vector $\vec q$ is defined relative to the appropriate  symmetry point, usually the $\Gamma$-point for  semi-conductors such as GaAs or GaIn.

It is however customary\cite{Winkler2003} to insist $\vec E$ is large only in the core region where the wave-functions are atomic like and it  follows\cite{Messiah1966}  $V_{\rm so} $ reduces to the familiar $v_{\rm so} =\xi(r)  \vec \sigma \cdot \vec \ell\, $ where $\xi (r) =\frac{e\hbar }{4m^2 c^2} \frac{1}{r} \frac{d V_1(r)}{dr} $ and where $V_1(r)$ is the Coulomb potential of the atomic core\cite{Winkler2003}. It is invariably  $v_{\rm so}$ that is accounted for in the current flavours of the DFT and in the usual theory of the Rashba effect as described, e.g., in Ref.~\cite{Winkler2003}. Using a number of different methods it is  well established\cite{Koseki2019}, 
 for isolated atoms, $v_{\rm so}$  accounts well for the spin-orbit splitting.

While the  wave-length $\lambda_c$ is very small  when compare with the Bohr radius $a_0$ or 
 the atomic radius of heavy elements, e.g.,  135pm for Au, it is not so small when compared with  the radius of the e.g., Au {\it inner\/} 1s orbital. This  goes as $r_1 = \frac{a_0}{Z}< 1$pm, where $a_0 \approx 53$pm is the Bohr radius and $Z=79$, for Au, the atomic number. If instead $a_0$ is replaced by the actual atomic radius of 135pm, it remains the case that $\frac{135}{Z}\sim 2$pm and fully comparable with $\lambda_c$.
 The speed of such a core $s$-electron $v_1 \sim (Z\alpha)c$ where $\alpha \approx \frac{1}{137}$ is the fine structure constant, and so
 for Au, $v_1 \sim 0.6$c. The electric field seen by this orbital $E_1\approx  k_e\frac{eZ^3}{{a_0}^2}$, where $k_e = 8.99 \times 10^9$Nm$^2$/C$^2$ is enormous as indicated in the text and below. The magnitude of $E_{\rm so} \sim  \frac{ 1}{8\pi} \alpha  (Z\alpha)  k_e\frac{eZ^3}{{a_0}} \sim \frac{ 1}{8\pi}  (Z\alpha)^2 Z^2 k_e\frac{e}{a_0}$, where $k_e\frac{e}{a_0} \sim 10$eV, and is much greater than the binding energy of hydrogen. 
 Landau and Lifshitz\cite{Landau1991} argue the {\it outer\/}, 6s for Au, orbitals have a weight $\frac{1}{Z^2}$ in the core region. Even then the 
  question not why is the spin-orbit coupling so big in the solid state, e.g., for Au, but rather why is it so small. 
  
In more detail, Landau and Lifshitz\cite{Landau1991} observe the WKB radial wave-function in the inner region goes as
$
|\psi| \sim \frac{1}{r\sqrt{p}}.
$
The logic is that $|\psi| = \frac{u}{r}$ where $u(r)$ is  non-singular. The $1/\sqrt{p}$ is the usual per-factor in the WKB wave-function. This is $|\psi|$, so the $e^{ip r/\hbar}$ does not appear. While this WKB approximation has well known problems at the origin  the above version  should couple smoothly with the exact solution for $s$-waves near the origin and should be good enough for the estimation being made here. In the WKB approximation, for energy $E$, the value of $p$ is given by
$
\frac{p^2}{2m} -| U(r)| = E
$
and near the core $E$ is negligible compared with  the potential $| U(r)| $.   Without worrying about the constants of proportionality:
$
p \sim \sqrt{| U(r)|}
$
and for the lowest energy orbitals $U(r) \approx \frac{Ze^2}{r} \sim \frac{Z^2e^2}{a_0} $
$
\sqrt{p} \sim | V(r)|^{1/4} \sim  \frac{Z^{1/2} }{{a_0}^{1/4} } .
$
It follows:
$$
|\psi| \sim \frac{1}{r\sqrt{p}} \sim  \frac{Z}{a_0\sqrt{p}} \sim \frac{Z}{Z^{1/2}} \sim Z^{1/2}
$$
The factor of $a_0$ is dropped since $|\psi| $ must be dimensionless. Then for an $s$-wave the wave-function $|\psi| $ should be a constant for $r < a_0/Z$ and so the probability of finding the wave-function in the lowest $s$-orbital is proportional to the volume $v_c = \frac{4\pi}{3}r^3\sim \frac{1}{Z^3}$ (not a shell) and the probability 
$
\sim |\psi|^2 v_c \sim Z\frac{1}{Z^3} \to \frac{1}{Z^2} 
$
again not keeping track of the dimension full $a_0$.  The estimate for the magnitude  $E_{so}$ of the spin-orbit coupling is then based upon 
$
E_{so} = \frac{e\hbar^2 }{4m^2 c^2} Ze^2 \langle \frac{1}{r^3}\rangle \to \frac{1}{4}\ Z^2 (\alpha Z)^2 \frac{e^2}{{a_0}} \to \frac{1}{4}\ Z^2\frac{e^2}{{a_0}}
$
dividing by the factor of $Z^2$. This 
Landau and Lifshitz\cite{Landau1991} estimate is based upon 
$
\langle \frac{1}{r^3} \rangle \sim \frac{Z^3}{ {a_0}^3} .
$
A better estimate for other than $s$-states is based upon  the  result\cite{radialint} 
$
\langle \frac{1}{r^3} \rangle =\frac{2}{n^3\ell(\ell+1) (2\ell+1) {a_0}^3}.
$
With the $Z^2$ correction  
$$
E_{so} = \frac{1}{2}\frac{1}{n(\ell + \frac{1}{2})(\ell+1)} (\alpha Z)^2 |E_{n}| 
$$
 where 
$
E_n =  - \frac{e^2}{2{a_0}} \frac{1}{n^2} 
$
are the hydrogen energy levels. This is the estimate reproduced in the text. That for a spherical potential $V(r)$, the spin-orbit coupling requires a non-zero $\ell$ and hence the mixing is 
 with core electrons with a finite $n >1$, i.e., a minimum $n=2$ and $\ell =1$. This is clearly important and leads to the moderate values of $E_{so}$ observed in solids. 

 Turning more specifically to the estimation of the Rashba spin-orbit coupling. Again with Au in mind, in the core region, using the above methods  $\langle E\rangle \approx  1.2  \times 
10^{16}$V/m while for $\ell =1$ the  (tangential) velocity $v \sim 0.15c$. In the context of electrical gating, it is hard to envisage an electric field much greater than $10^{11}$V/m and, due to the weighting factor of $1/Z^2$,  the Rashba magnetic field $B_R$ due to the core  is very much negligible, while for the outer electrons the $\ell = 1$ tangential velocity is very much too small. 

Consider instead the missing radial velocity. The first estimate is  the momentum associated with the Bohr radius. Here  $\hbar k = \hbar/ a_0 = (1.05 \times 10^{-34})/ (0.5 \times 10^{-10}) \approx  2.0 \times 10^{-24}$ and so the velocity is $v = \hbar k /m= 2.0 \times 10^{-24}/  (9.1\times 10^{-31}) = 2 \times 10^6$m/s and a typical Fermi velocity, reflecting a crystal momentum. However 
for a n$s$-orbital there are $n-1$ radial nodes implying $2n-1$ nodes across the atomic diameter. 
If, as for Au,  $n =6$  then $v \approx 1.2 \times 10^7$m/s or approx $\frac{c}{25}$. 

The effect of the crystal potential is very important. The  Au-Au bond length is $a = 2.84$\AA\ and, given a simple cubic structure assumed in the text, it is this and not the actual lattice parameter that should be used for $a$. It should really  be less than this since the interstitial region has been shrunk to zero in the derivation of the delta function model. This notwithstanding for  $a = 2.84$\AA\  it follows, with $G_n = \frac{2\pi}{a} n$, the momentum $\hbar G_n \approx  \frac{2\pi}{a} n 
= 2.2 n \times 10^{10} $ giving  $v = \hbar k /m\approx  n\frac{c}{118}$. So for the Kronig-Penny model with $n_x=11$ appropriate to the {\it outer\/} Au electrons $v \approx \frac{c}{10}$ and already comparable to the $\ell=1$ tangential velocity {\it in the core}. If  interstitial region is shrunk to zero, the minimum $a \sim 1$\AA\ corresponding to two Bohr radial touching each other, i.e., $a \approx 1$\AA\ and gives an upper limit $v \approx \frac{c}{3.5}$. Thus while it is impossible to get an experimentally significant Rashba coupling from the core region, it is possible to have a useful contribution from the outer electron {\it radial velocity\/} of heavy elements due to the large principal quantum number $n$. This is the theme taken up in the text.

{\it Zitterbewegung  oscillations.} 
For a  single free  electron described by Dirac's equation, the Compton wave-length $\lambda_c \approx 2$pm and is about five orders of magnitude smaller the experimentally determined value in some solids, e.g, for InAs\cite{Iwasaki2017}  $\lambda_c \sim 100$nm. 
Zitterbewegung (ZB) oscillations are a basic measurement that should set the scale for the key parameter $\lambda_c$. It is often suggested\cite{Manchon2015} that the large value of spin-orbit effect, in general, and specifically the value of the frequency associated with the ZB oscillations, reflect the fact that the mass gap $mc^2 \approx 0.5$MeV is enormously larger than the spin-orbit band gap $\lesssim 1$eV in the above and many other semi-conductors. While this argument may be correct in this context it does not explain the similarly large magnitude of the Rashba coupling.

{\it The central potential spin-orbit coupling as a perturbation
} 
It is important to understand the origin of the Rashba spin-orbit coupling that arises from 
 the more usual reduced spin-orbit coupling $v_{\rm so} = \xi (r) \vec \sigma \cdot \vec \ell$. In the further reduction of this to  traditional Rashba and/or Dresselhaus form, $ \vec p$ in  the original $V_{\rm so} $ gets replaced by the crystal momentum $\vec q$. The current exercise has shown that for a simple periodic model this reduction is not valid. It is a nicety of the model examined in the text  that $\vec E $ is the ``applied" electric field, i.e., the angular momentum $\vec E \times \vec p$ has no intrinsic contribution because the potential within each cell, or effective atom, has no gradient. A small radial potential $V(r)$ is now added, as a perturbation,
in order to examine the role of $v_{\rm so}$. 
A typical perturbative contribution\cite{Wang1996} to the magnetic anisotropy, due to $v_{\rm so}$, has the splitting between, e.g., the t$_{2g}$ to e$_g$ levels as the energy denominator whereas, for the model in the text,
 this is the exchange $J_0 S$. The quantum numbers that couple are different and hence gaps that might appear correspond to different band crossings. This is of some considerable experimental importance.

Explicitly $ \xi (r) \vec \sigma \cdot (\vec r \times \vec p)  = \xi (r)\sigma_z ( x p_y - yp_x) + (\xi z)(\sigma_y p_x - \sigma_x p_y) + \xi(\sigma_x y - \sigma_y x)p_z$.  In order to make calculations more tractable, take $\xi$ to be a constant. It has been implicitly  assumed the wave-function for the $z$-direction is node-less and that there is symmetric wave-function  $\psi_0(z)$ corresponding to the lowest energy. The quantity $p_z\psi_0(z)$ has a single node and is orthogonal to this state. The energy to the first excited state is large compared to other excitation energies and there are therefore no relevant matrix elements of $\xi(\sigma_x y - \sigma_y x)p_z$ in perturbation theory. If the expectation value of $z$ is finite, i.e., if inversion symmetry is broken, the term $ (\xi \langle z \rangle )(\sigma_y p_x - \sigma_x p_y) $ screens  the direct Rashba effect. The ``ionic" polarisation  $- e\langle z \rangle $ is  due to an internal effective field $E_i$ that is already accounted for in the direct Rashba term. If  the dielectric constant $\epsilon$ is large, the effect of the polarisation is to reduce $E_i \to E_i/\epsilon$. A large pseudo Rashba effect, reflecting $ (\xi \langle z \rangle )(\sigma_y p_x - \sigma_x p_y) $ and a large  polarisation, will actually cancel the direct Rashba and/or Dresselhaus effects. It is therefore highly problematic if a DFT code accounts for  $ \xi \vec \sigma \cdot (\vec r \times \vec p) $ but not the direct Rashba effect treated here.

It remains to account for the part $\xi \sigma_z \ell_z = \xi  \sigma_z(x p_y - yp_x)$ that involves the matrix elements of  $ \ell_z = x p_y - yp_x$.   Consider the matrix elements of $x p_y $. The action of $p_y = - i \hbar \partial_y$ is to add an extra node to $\psi_{k_y} (y)$. As a consequence there are only off-diagonal matrix elements for the $\psi_{k_y} (y)$ part of the factored wave-function. The periodic  $x = \frac{a}{2\pi} i  \sum_{n_x=1}^\infty \frac{1}{s_x} (-1)^{s}\left(e^{iG_{n_x}x} - e^{iG_{-n_x}x} \right)$ when written in terms of the reciprocal lattice vectors $G_{n_x}$. It is again the case that the factor of $e^{iG_{n_x}x} $ in the expansion imply there are also only off-diagonal matrix elements for the $\psi_{k_x} (x)$ part of the wave-function. Perhaps more direct is the observation that a factor of $x$ turns an odd to an even function or the inverse and also adds a node. The matrix elements of  $\xi \sigma_z \ell_z $ connect to states that are doubly off-diagonal. For the tight binding bands it is implies that both $n_x$ and $n_y$ differ by at least unity. While possible, because of the approximation that $\xi$ is a constant, it is not particularly useful to continue with the evaluation of the matrix elements. It is however important to understand the magnitude of the magnetic anisotropy. The matrix elements of, e.g., $x p_y \sim \hbar a \frac{2\pi}{a} s_y$ and those of $(x p_y - yp_x)$ squared $\sim 4 \hbar^2 \pi^2 ({s_x}^2 + {s_y}^2)$. Unlike the Rashba coupling here these matrix elements are between states that differ by their orbital quantum numbers $s_x$ and $s_y$ for the tight-binding regime of the model. These states have energy $E_0({s_x}^2 + {s_y}^2)$ for which $s_x$ and $s_y$ change by at least unity. The anisotropy field $B_A$ is therefore of order $4 \xi^2  \hbar^2 \pi^2  \sqrt{{s_x}^2 + {s_y}^2}$. Then  
since $\xi \sim E/a$, where $E\sim e/a^2$ is the magnitude of the radial electrical field at the average position of the electrons, it follows $B_A \sim  4 \pi^2   \sqrt{{s_x}^2 + {s_y}^2} \frac{\alpha^2 }{a^2}E^2$. In the present approach, the Rashba field $B_R$ rather involves matrix elements between states that have $\sigma_z =\pm 1$ and an energy difference $J_0S$ independent of $s_x$ and $s_y$. This leads to a $B_R \sim  4 \pi^2  ( {s_x}^2 + {s_y}^2) \frac{\alpha^2 }{a^2}E^2$ with a stronger dependence on $s_x$ and $s_y$. It is again emphasised that degenerate states with all $s_x = s_y = s_z =s$, and $s$ odd, correspond to higher ``s-bands'' with $s = 2n-1$ where $n$ is the principal quantum number and that $( {s_x}^2 + {s_y}^2) \to ( {s_x}^2 + {s_y}^2 +{s_z}^2)$ for three dimensional atoms imbedded in two dimensions. For heavy elements such as Au, this increases the Rashba coupling by two orders of magnitude as compared with free electrons and an order of magnitude relative to the $\xi(r)  \vec \sigma \cdot \vec \ell $ spin orbit terms, see the discussion in the text.

{\it The Hohenberg-Kohn theorem.
} 
As Berry\cite{Berry1984}  has pointed out, in the context of adiabatic evolution of a spin dependent Hamiltonian,  the spin dependent phase of a wave-function serves an essential purpose. However since the Hamiltonian is time dependent this observation does not have direct repercussions on the Hohenberg-Kohn theorem that insists the ground state energy is a functional of only the density, i.e., depends only on the modulus of the wave-function.  It is however well known\cite{Ye2001a,Ye2001b,Ye2001c}, in the context of the spin-Hall effect, that spin texture, i.e., the existence of non-collinear magnetisation, introduces a spin ``flux" $\Phi_s$ with non-trivial consequences even for Hamiltonians that are time independent. More relevant here is  that the energy of electrons on a ring with a certain configuration of static magnetic fields, i.e., the Stern problem\cite{{Shankar1994}}  has a time independent Hamiltonian. There is  a non-degenerate ground state energy that depends on the spin Berry phase $\phi_s = \frac{\hbar}{e}\Phi_s$ through the boundary conditions\cite{{Shankar1994}}. For the free electron Rashba problem there is a similar configuration of static magnetic fields and  this result applies. 
However it is easy to argue the boundary conditions do not effect the bulk density of states and hence the energy in the thermodynamics limit.

In the presence of a periodic potential it has been shown the Kronig-Penney  problem  reduces to finding the roots of 
$$
 \frac{1}{V} =
             \frac{1}{2} \sum_\pm \sum_{n_x}
            %e^{\pm i\frac{\pi}{4}(\cos \theta_{G_{n_x}} \sigma_z + \sin \theta_{G_{n_x}} \sigma_y)}  
            (1  \pm   e^{ -i \frac{\pi}{2} (\cos \theta_{G_{n_x}} +\sigma_z)\sigma_z - \sin \theta_{G_{n_x}} \sigma_y) } )
            {\cal G}^{\pm}_{G_{n_x}}
%\tag{S49}\label{S49}
$$  
where the $ {\cal G}^{\pm}_{G_{n_x}}$ are real Green's functions that reflect the dynamic phase. It has also been seen that the Berry phase factor 
$%e^{\pm i\frac{\pi}{4}(\cos \theta_{G_{n_x}} \sigma_z + \sin \theta_{G_{n_x}} \sigma_y)} 
e^{ -i \frac{\pi}{2} (\cos \theta_{G_{n_x}} +\sigma_z)\sigma_z - \sin \theta_{G_{n_x}} \sigma_y) }
$ is a reflection of a real space spin off-diagonal Berry connection $\vec A_s(\vec r)$. Now the spin Berry phase appears not only in the boundary conditions but also in the reflection amplitudes. 
It is possible to define an effective magnetic field, i.e., the Berry curvature, that is proportional to   the commutators $[{A_s}_i , {A_s}_j ]$ of the components ${A_s}_i$. This is distinct from the Rashba field $\vec B_R$, but is the flux of this field that modulates the spin interference effects. 
Since the angles $ \theta_{G_{n_x}} $ are self-consistently determined $\vec A_s(\vec r)$ cannot be considered as an external potential that might be accommodated within an extended the Hohenberg-Kohn theorem \cite{Hohenberg1964,Vignale1988}.